\documentclass[12pt,a4paper]{article}
\usepackage{graphicx,amssymb,amsmath,color,scalefnt}
\usepackage{multirow, makecell}
\usepackage{soul}
\usepackage{chngpage}
\usepackage{geometry}
\usepackage[dvipsnames]{xcolor}
\usepackage{rotating}
\usepackage[normalem]{ulem}
\usepackage{jheppub}

\def\TeV{\ifmmode {\mathrm{Te\kern -0.1em V}}\else
                   \textrm{Te\kern -0.1em V}\fi\,}%
\def\GeV{\ifmmode {\mathrm{Ge\kern -0.1em V}}\else
                   \textrm{Ge\kern -0.1em V}\fi\,}%
\def\MeV{\ifmmode {\mathrm{Me\kern -0.1em V}}\else
                   \textrm{Me\kern -0.1em V}\fi\,}%
\def\keV{\ifmmode {\mathrm{ke\kern -0.1em V}}\else
                   \textrm{ke\kern -0.1em V}\fi\,}%
\def\eV{\ifmmode  {\mathrm{e\kern -0.1em V}}\else
                   \textrm{e\kern -0.1em V}\fi\,}%
\let\tev=\TeV

\def\ifb{\mbox{fb$^{-1}$}}
\def\ipb{\mbox{pb$^{-1}$}}

\definecolor{Purple}{rgb}{0.6, 0.4, 0.8}

\newcommand{\com}[1]{{ \color{red} #1}}
\newcommand{\myComment}[1]{}
\newcommand{\HEPfit}{\texttt{HEPfit }}

\DeclareMathAlphabet{\mathsfit}{\encodingdefault}{\sfdefault}{m}{sl}
\newcommand{\ges}[1]{\mathsfit{#1}}

\newcommand{\FDF}[1][]{\varphi^\dagger #1\!\overleftrightarrow{D}\!_\mu\varphi}
\newcommand{\FDFI}[1][]{\varphi^\dagger #1\!\overleftrightarrow{D}^I\!\!\!_\mu\:\varphi}

\begin{document}

\title{The top quark electro-weak couplings after LHC Run~2}

\author[a]{V\'ictor Miralles,}
\emailAdd{victor.miralles@ific.uv.es} 
\author[a]{Marcos Miralles L\'opez,}
\emailAdd{marcos.miralles@ific.uv.es}
\author[a]{Mar\'ia Moreno Ll\'acer,}
\emailAdd{maria.moreno@ific.uv.es}
\author[a,b]{Ana Peñuelas,} 
\emailAdd{ana.penuales@ific.uv.es}
\author[a]{Martin Perell\'o} 
\emailAdd{martin.perello@ific.uv.es}
\author[a]{and Marcel Vos}
\emailAdd{marcel.vos@ific.uv.es}

\affiliation[a]{Instituto de F\'isica Corpuscular (IFIC, Universitat de Val\`encia and CSIC),\\C/ Catedr\'atico Jos\'e Beltr\'an 2, E-46980 Paterna, Valencia, Spain}
\affiliation[b]{PRISMA+ Cluster of Excellence, Johannes Gutenberg University Mainz,\\Staudingerweg 9, 55128 Mainz, Germany}

\abstract{
Recent measurements at the Large Hadron Collider allow for a robust and precise characterisation of the electro-weak interactions of the top quark. We present the results of a global analysis at next-to-leading order precision including LHC, LEP/SLD and Tevatron data in the framework of the Standard Model Effective Field Theory. We include a careful analysis of the impact of correlations among measurements, as well as of the uncertainties in the Effective Field Theory setup itself. We find remarkably robust global fit results, with central values in good agreement with the Standard Model prediction, and 95\% probability bounds on Wilson coefficients that range from $\pm 0.35$ to $\pm 8$ \TeV$^{-2}$. This result represents a considerable improvement over previous studies, thanks to the addition of differential cross-section measurements in associated production processes of top quarks and neutral gauge bosons.
}

\keywords{top quark, electro-weak couplings, effective field theory}






\preprint{{\raggedleft IFIC/21-28 \par}}

\setcounter{tocdepth}{1}

\maketitle

\section{Introduction}

The Standard Model (SM) of particle physics offers an accurate description of the interactions of elementary particles. A large number of precise measurements at high-energy colliders test the theory at the \tev{} scale. The Standard Model Effective Field Theory (SMEFT) extends the theory with higher dimension operators compatible with the gauge symmetry and provides a systematic framework to analyze the wealth of data in terms of potential new physics.

The top quark, the most massive particle of the SM, has tight connections to the Higgs boson and plays a prominent role in many extensions of the SM. In particular, the top and bottom quark electro-weak couplings can be modified in composite Higgs/extra dimension models and scenarios with vector-like quarks~\cite{Richard:2014upa, Durieux:2018ekg,Dawson:2020oco}. Its properties make it particularly suited for experimental investigation and a rich experimental programme has developed around the top quark since its discovery in 1995~\cite{PhysRevLett.74.2626,D0:1995jca}. The Tevatron experiments could characterize the top quark interactions with the gluon, through the QCD production mechanism, and with the $W$~boson, through the electro-weak decay $t\rightarrow Wb$ and single top quark production processes. The Large Hadron Collider (LHC) experiments provide new measurements of these traditional probes with increased precision and in an extended kinematic regime, and have observed rare associated production processes that provide a direct probe for the couplings of the top quark with the photon, the $Z$~boson and the Higgs boson.

Compared to previous Effective Field Theory (EFT) analyses of the top quark sector~\cite{Durieux:2019rbz,Brivio:2019ius, Hartland:2019bjb,Buckley:2015nca,Buckley:2015lku}, we extend the set of measurements considerably. We incorporate recent differential measurements in $pp\rightarrow t\bar{t} \gamma$ and $pp \rightarrow t\bar{t}Z$ production, and measurements of the $pp \rightarrow t\bar{t}W$, $pp \rightarrow t\bar{t}H$, $pp \rightarrow t Z q$ and $pp \rightarrow t \gamma q$ inclusive cross-sections. The precise measurements of the $Z\rightarrow b\bar{b}$ vertex at LEP and SLD~\cite{ALEPH:2005ab} remain very relevant and are included, as well as legacy Tevatron measurements.

In this paper we use this large set of measurements at high-energy colliders to provide a complete characterisation of the electro-weak couplings of the top quark. We derive bounds on eight Wilson coefficients of the relevant dimension-six two-fermion operators. An important effort is made to ensure that the results are robust. Importantly, we estimate the correlations between measurements and evaluate their impact on the fit results. The main limitations of the SMEFT analysis, due to the truncation of the expansion at dimension-six and due to the suppression of certain degrees of freedom, are also addressed in detail. As the main result of this paper, we present robust limits on the effects of high-scale phenomena beyond the SM on the top quark electro-weak couplings that include the most recent LHC Run~2 results.

Several groups have performed combined EFT fits of the top, Higgs and electro-weak sectors~\cite{Ellis:2020unq,Ethier:2021bye}, while others have explored the interplay between top quark physics and the flavour factories~\cite{Bruggisser:2021duo, Bissmann:2020mfi, Bissmann:2019gfc, Cirigliano:2016nyn, Alioli:2017ces}. In addition, three recent studies explore events involving top quarks and $Z$~bosons to further constrain some electro-weak couplings~\cite{CMS:2021rvz,Keaveney:2021dfa,Ravina:2021kpr}.

\section{Effective Field Theory framework}

Effects of new physics in the couplings of the top quarks can be described as effective
interactions of SM particles at energies below a new physics matching scale $\Lambda$. These effective interactions can be parameterised in terms of a set of Wilson coefficients $C_i$ of dimension-six operators $O_i$ in
the effective Lagrangian,

\begin{equation}
\mathcal{L}_\text{eff} = \mathcal{L}_\text{SM} +   \left( \frac{1}{\Lambda^2} \sum_i \mathcal{C}_i O_i + \text{h.c.} \right)  + \mathcal{O}\left(\Lambda^{-4} \right) ,
\end{equation}

\noindent
where the sum runs over a total of ten operators that involve top quarks, as described below, and which can be interpreted in terms of new physics mediators. This EFT preserves the local and gauge symmetries of the SM, and operators with odd dimension are omitted since they will violate baryon or lepton number.

The SMEFT predictions of a physical quantity, $X$, can be written in this form:
\begin{equation}
\begin{array}{cc}
X = X_{SM} + \sum_i \frac{\mathcal{C}_i}{\Lambda^2} X_i^{(1)} + \sum_{ij} \frac{\mathcal{C}_i \mathcal{C}_j}{\Lambda^4} X_{ij}^{(2)}+ \mathcal{O}\left(\Lambda^{-4} \right)  = \\
X_{SM} \times  \left( 1 + \sum_i\frac{\mathcal{C}_i}{\Lambda^2} \frac{X_i^{(1)}}{X_{SM}} + \sum_{ij} \frac{\mathcal{C}_i \mathcal{C}_j}{\Lambda^4} \frac{X_{ij}^{(2)}}{X_{SM}} \right) + \mathcal{O}\left(\Lambda^{-4} \right). \end{array}
\label{eq:XS_dep}
\end{equation}

\noindent
This expression contains the SM part, the interferences of the effective dimension-six operators with the SM (referred to as ``linear terms'') which are proportional to $\mathcal C_i /\Lambda^2$, and 
``quadratic terms'' proportional to $\mathcal C_i\mathcal C_j / \Lambda^4$. The effects of dimension-eight operators and the so-called ``double insertions'' terms of dimension-six operators that contribute at $\Lambda^{-4}$ order are not included. To assess the validity of the EFT expansion, we compare fits with and without $\Lambda^{-4}$ terms, as recommended by the LHC Top Working Group~\cite{AguilarSaavedra:2018nen}, and discuss the robustness of the bounds in detail in Sections~\ref{sec:globalbounds} and \ref{sec:robust}.

The number of operators involved in the most general EFT description is prohibitive for an analysis of this type. Therefore, here we select a subset of operators that provides an adequate basis for a study of new physics effects in the top and bottom quark electro-weak couplings. We focus on the two-fermion operators that affect top and bottom quark interactions with the vector, tensor or scalar Lorentz structures listed below, and we use the Warsaw basis~\cite{Grzadkowski:2010es} (see also Refs.~\cite{AguilarSaavedra:2008zc,Zhang:2010dr}) following the conventions proposed by the LHC Top Working Group~\cite{AguilarSaavedra:2018nen}:
\begin{equation}
\begin{array}{@{}rlcc@{}}
    \underline{O_{\varphi Q}^1}
		&  \equiv 
		\frac{1}{2} \left(\bar{\ges q} \gamma^\mu \ges q \right) \left(\FDF[i] \right) 	,\\
	O_{\varphi Q}^3
		&\equiv 
		\frac{1}{2} \left(\bar{\ges q} \tau^I \gamma^\mu \ges q \right) \left(\FDFI[i] \right) ,\\
	O_{\varphi Q}^-
		&\equiv 
		O_{\varphi Q}^1 - O_{\varphi Q}^3
	,\\
	O_{\varphi u}
		&\equiv \frac{1}{2}
		\left(\bar{\ges u}\gamma^\mu \ges u\right)\left(\FDF[i]\right)
		,\\
	O_{\varphi d}
		&\equiv \frac{1}{2}
		\left( \bar{\ges d}\gamma^\mu \ges d \right)\left( \FDF[i]\right)
        ,\\
    O_{\varphi ud} 
        &\equiv \frac{1}{2}
        	\left(\bar{\ges u} \gamma^\mu \ges d \right) 	\left(\varphi^T \epsilon i D_\mu \varphi \right)
        ,
\end{array}
\quad
\begin{array}{@{}rlcc@{}}
	O_{uW}
		&\equiv 
		\left(\bar{\ges q}\tau^I\sigma^{\mu\nu} \ges u  \right)\left(\epsilon\varphi^* W_{\mu\nu}^I  \right)
	,\\
        O_{dW}
        &\equiv	\left(
		\bar{\ges q}\tau^I\sigma^{\mu\nu} \ges d  \right) \left(	\varphi W_{\mu\nu}^I  \right)
	,\\
	\underline{O_{uB}} & \equiv	\left(\bar{\ges q}\sigma^{\mu\nu} \ges u \right)
	 \left( \epsilon\varphi^* B_{\mu\nu} \right)  ,  \\
	 \underline{O_{dB}}
		&\equiv 	\left( \bar{\ges q}\sigma^{\mu\nu} \ges d \right) \left(\varphi  B_{\mu\nu} \right), \\
	O_{uZ}
	&\equiv - \sin \theta_W
	O_{uB} + \cos \theta_W O_{uW}
		,\\
    O_{dZ}
		&\equiv - \sin \theta_W
		O_{dB} + \cos \theta_W O_{dW}
		,\\
	O_{u\varphi}
		&\equiv  
		\left( \bar{\ges q} \ges u \right)
		\left( \epsilon\varphi^* \; \varphi^\dagger\varphi \right)
	,\\
	O_{d\varphi}
		&\equiv  
		\left( \bar{\ges q} \ges d \right)
		\left( \epsilon\varphi^* \; \varphi^\dagger\varphi \right).
\end{array}
\label{eq:op_2q}
\end{equation}


The underlined operators $O_{\varphi Q}^1, O_{uB}$ and $O_{dB}$ are not used directly, but are present in the linear combinations $O_{\varphi Q}^-$, $O_{uZ}$ and $O_{dZ}$. The operators $O_{\varphi Q}^1$ and $O_{\varphi Q}^3$ modify the left-handed couplings of the $Z$~boson to down-type and up-type quarks. Two further operators $O_{\varphi u}$ ($O_{\varphi d}$) modify the right-handed couplings of the $Z$~boson to up-type (down-type) quarks. The electro-weak dipole operators labeled $O_{uW}$ ($O_{dW}$) and $O_{uZ}$ ($O_{dZ}$) give rise to tensor couplings of the photon and $Z$~boson to the up-type (down-type) quarks and induce an anomalous dipole moment.
The $O^3_{\varphi Q}$ and $O_{uW}$ operators also modify the charged-current interactions of the up-type quarks with a $W$~boson and left-handed down-type quark, while $O_{\varphi ud}$ and $O_{dW}$ give rise to interactions between the up-type quarks with a $W$~boson and right-handed down-type quark.
Finally, $O_{u\varphi}$ and $O_{d\varphi}$ modify the Yukawa couplings of up-type and down-type quarks.

Operators that modify only the bottom quark electro-weak couplings, $O_{\varphi d}$ and $O_{dZ}$, are taken into account in the fit since we include measurements from electro-weak precision data, but limits on their coefficients are not reported in this paper since the obtained values are not competitive enough using only the observables considered in our fit. The operator that shifts the bottom quark Yukawa, $O_{d\varphi}$, is not included in the fit, as the measurements considered here are not sensitive to it. 

For the fits without (with) $\Lambda^{-4}$ terms, we report results for six (eight) operator coefficients that characterize the top quark couplings to the $Z$~boson, the photon and the Higgs boson and the charged-current $tWb$ vertex. Substituting $u$ for $t$ and $d$ for $b$, the list reads:
\begin{displaymath}
 C_{\varphi Q}^{-}, C_{\varphi Q}^{(3)}, C_{\varphi t}, C_{tW}, C_{tZ}, C_{t \varphi}, (C_{\varphi tb}, C_{bW}).
\end{displaymath}

This choice of basis omits several degrees of freedom.  The CP-violating imaginary parts of those operators are left for future work. The four-fermion operators with two light quarks and two heavy quarks ($qqQQ$), and the two-fermion operator $O_{tG} = \left(\bar{q} \sigma^{\mu \nu} T^{A} u \right)\left(\epsilon \varphi^* G_{\mu \nu}^{A} \right)$, that modifies the $t\bar{t}g$ vertex, are not included in the baseline fit, but their impact is studied in Section~\ref{sec:robust}. Finally, four-fermion operators with two leptons and two heavy quarks ($\ell \ell QQ$) are not included. These can be constrained with carefully constructed measurements at the LHC~\cite{Sirunyan:2020tqm} and through loop effects in B-factories~\cite{Bissmann:2020mfi,Bruggisser:2021duo}. Very strong bounds can be derived at a future lepton collider~\cite{Durieux:2019rbz}.

\section{Measurements}

In this work, we include for the first time the differential cross-section measurements of $t\bar{t}Z$ and $t\bar{t}\gamma$ processes (sensitive to top quark neutral-current interactions).  
Inclusive cross-section measurements are considered for $t\bar{t}W$ and $t\bar{t}H$ production, single top production in the t-channel ($tq$), associated production ($tW$) and s-channel ($t\bar{b}$), and single top quark production in association with $Z$ boson or a photon ($tZq$ and $t\gamma q$). We also include $W$~boson helicity fractions (sensitive to charged-current interactions). Finally, measurements of $R_b$ and $A_{FBLR}^{b\bar{b}}$ in bottom quark pair production at the $Z$-pole from the LEP and SLD experiments (sensitive to operators that affect the left-handed couplings of the top and bottom quarks) are included. The selected measurements are summarized in Table~\ref{tab:measurements}. Counting the bins in the differential cross-section measurements, the total number of observables ($n_{\text{obs}}$) is 30.

\begin{table*}[!ht]
\centering
\resizebox{\textwidth}{!}{  
\begin{tabular}{|l|c|c|c|c|c|c|}
\hline
Process & Observable & $\sqrt{s}$  & $\int \cal{L}$  & Experiment & SM  & Ref.\\ \hline
$p p \rightarrow t \bar{t} H+ tHq$ & $\sigma$ & 13 TeV & 140~\ifb & ATLAS & 
\cite{deFlorian:2016spz} &  \cite{ATLAS:2020qdt} \\ 
$p p \rightarrow t \bar{t} Z$ &  $d\sigma/dp_T^Z$ (7 bins) & 13 TeV &  140~\ifb  & ATLAS &
\cite{Broggio:2019ewu} & \cite{ATLAS:2020cxf} \\ 
$p p \rightarrow t \bar{t} \gamma$ & $d\sigma/dp_T^\gamma$ (11 bins) & 13 TeV & 140~\ifb & ATLAS &
\cite{Bevilacqua:2018woc,Bevilacqua:2018dny} & \cite{Aad:2020axn}  \\ 
$p p \rightarrow tZq$ & $\sigma$ & 13 TeV & 77.4~\ifb{}  & CMS &
 \cite{Sirunyan:2017nbr} & \cite{Sirunyan:2018zgs} \\ 
$p p \rightarrow t\gamma q$ & $\sigma$ & 13 TeV & 36~\ifb{}  & CMS &
 \cite{Sirunyan:2018bsr} & \cite{Sirunyan:2018bsr} \\
 $p p \rightarrow t \bar{t} W$ & $\sigma$ & 13 TeV & 36~\ifb  & CMS &
\cite{deFlorian:2016spz,Frederix:2017wme} &  \cite{Sirunyan:2017uzs} \\
 $p p \rightarrow t\bar{b}$ (s-ch) & $\sigma$ & 8~\tev{} & 20~\ifb{} & LHC &
\cite{Aliev:2010zk,Kant:2014oha} & \cite{Aaboud:2019pkc} \\ 
$p p \rightarrow tW$ & $\sigma$ & 8~\TeV & 20~\ifb{}  & LHC &
 \cite{Kidonakis:2010ux} &  \cite{Aaboud:2019pkc}   \\ 
$p p \rightarrow tq$ (t-ch) & $\sigma$ & 8~\tev{} & 20~\ifb{} & LHC &
\cite{Aliev:2010zk,Kant:2014oha} & \cite{Aaboud:2019pkc} \\ 
$t \rightarrow Wb $ & $F_0$, $F_L$  & 8~\tev{} & 20~\ifb{} & LHC &
\cite{PhysRevD.81.111503}  & \cite{Aad:2020jvx} \\
$p\bar{p} \rightarrow t\bar{b}$ (s-ch) & $\sigma$ & 1.96~\tev & 9.7~\ifb & Tevatron & \cite{PhysRevD.81.054028} & \cite{CDF:2014uma} \\
$e^{-} e^{+} \rightarrow b \bar{b} $ & $R_{b}$ ,  $A_{FBLR}^{bb}$ & $\sim$ 91~\GeV & 202.1~\ipb  & LEP/SLD &
 - & \cite{ALEPH:2005ab}  \\ \hline
\end{tabular}
}
\caption{Measurements included in the EFT fit of the top quark electro-weak sector. For each measurement, the process, the observable, the center-of-mass energy, the integrated luminosity and the experiment/collider are given. The last two columns list the references for the predictions and measurements that are included in the fit. LHC refers to the combination of ATLAS and CMS measurements. In a similar way, Tevatron refers to the combination of CDF and D0 results, and LEP/SLD to different experiments from those two accelerators.}
\label{tab:measurements}
\end{table*}

For LHC observables at $\sqrt{s}=$~13~\tev where no official combinations are available, measurements either from ATLAS or CMS experiment (but not both) are included. In case results are available for several final states depending on the decay mode of the top quarks or associated bosons, measurements are chosen to be (nearly) orthogonal. In addition, for observables that have been measured at multiple center-of-mass energies, only one measurement is included in order to avoid issues with correlations. The one exception is the $s$-channel single top quark production, where both Tevatron and LHC results are included since they have similar sensitivity and are independent data sets. In most cases, the latest Run~2 results at 13~$\tev$ are chosen, but for the measurements of the single top quark cross-sections and $W$~boson helicity fractions the combined (ATLAS+CMS) LHC 8~$\tev$ results are preferred. For two single top quark production channels, $pp\rightarrow t q$ and $pp\rightarrow t W$, the cross-sections have also been measured at 13 TeV (although no LHC combinations are available). We have checked that including such measurements at 13~TeV, in addition to those at 8~TeV, has a negligible impact in our limits. This check reinforces our strategy of including only the measurement at one center-of-mass energy.

Even among measurements of different processes, correlations might have a non-negligible effect, as experimental and modelling systematic uncertainties have common sources. The theory predictions can also be correlated, through the parton density functions and the similarity of the matrix elements of the several associated production processes. When available, we have included published experimental correlations, for instance for the LEP and SLD measurements~\cite{ALEPH:2005ab} and the $W$ boson helicity fractions~\cite{Aad:2020jvx}. In the case of  $t\bar{t}\gamma$ and $t\bar{t}Z$ differential cross-section measurements~\cite{Aad:2020axn, ATLAS:2020cxf} as function of the gauge boson transverse momentum, the correlations among bins provided by the experiments are also included. For the remaining observables, we have defined an ansatz correlation matrix to test the robustness of the fit results (see Section~\ref{sec:robust}).

Several processes deserve some further discussion. The $t\bar{t}\gamma$ cross-section measurement receives contributions from the $tW\gamma$ process and from $t\bar{t}$ events, where one of the top quark decay products (\emph{i.e.} the $b$~quark or $W$~boson or its decay products) emits a photon. Events without a top-photon vertex dilute the sensitivity to the EFT operator coefficients~\cite{Bevilacqua:2019quz}. The fiducial region definition and the differential analysis in the ATLAS measurement~\cite{Aad:2020axn} mitigate their impact, as the photons emitted by decay products tend to be close to jets and are predominantly soft. We account for $tW\gamma$ and photons from top decay products fully in the SM prediction, but ignore these contributions in the EFT parameterisation.

Similarly, the $t\bar{t}H$ cross-section measurement includes also $tH$ (due to the difficulty to disentangle the two processes experimentally). The latest ATLAS combination of several final states has been used~\cite{ATLAS:2020qdt}. The $tH$ contribution is included in the SM prediction and in the EFT parameterisation.

The $t\bar{t}W$ measurement from the CMS experiment~\cite{Sirunyan:2017uzs} targets the multi-lepton final state and is inclusive in the number of jets. Thus, the parameterisation considers the $pp \rightarrow t\bar{t}W$ process at NLO in QCD and additionally explicitly takes into account the contribution of the electro-weak $t\bar{t}Wq$ production, as done in Ref.~\cite{Banelli:2020iau}. The latter process includes $tW\rightarrow tW$ scattering diagrams which have a strong sensitivity to $C_{\varphi t}$ operator~\cite{Dror:2015nkp}. 

\section{Fit setup}
\label{subsec:parameterisation}

The dependence of the observables included in the fit on the Wilson coefficients, that is the $X_i^{(1)}/X_{SM}$ and
$X_{ij}^{(2)}/X_{SM}$ terms in Eq.~\ref{eq:XS_dep}, is evaluated with the Monte Carlo generator \texttt{MadGraph5\char`_aMC@NLO} v.2.7.0~\cite{Alwall:2014hca} using a fixed order calculation except for $t\to Wb$ decay for which the analytical calculation at NLO is used~\cite{Zhang:2014rja}. Two UFO models are employed. The \texttt{SMEFT@NLO}\footnote{Either \texttt{SMEFTatNLO\char`_U2\char`_2\char`_U3\char`_3\char`_cG\char`_4F\char`_LO\char`_UFO} or v1.0.1 are used.} model~\cite{Degrande:2020evl} is used for most of the operators and the predictions are derived at NLO in QCD, except for the observables from LEP, SLD and Tevatron colliders which are parameterised at LO. The parameterisation for the bottom-quark operators ($\mathcal{O}_{bW}$, $\mathcal{O}_{\varphi tb}$, $\mathcal{O}_{bZ}$ and $\mathcal{O}_{\varphi b}$) are obtained with the \texttt{TEFT\char`_EW} model~\cite{Bylund:2016phk}. Both linear and quadratic terms are considered, as well as the interference between the different operators.

The \texttt{SMEFT@NLO} model uses the $m_W,m_Z,G_F$ input electro-weak scheme while the \texttt{TEFT\char`_EW} model uses the $\alpha_{EM}^{-1}(m_Z),m_Z,G_F$ scheme:

\begin{equation*}
    \alpha_{EM}^{-1}(m_Z) = 127.95, \qquad G_F = 1.16637\times 10^{-5}~\text{GeV}^{-2},    
\end{equation*}
\begin{equation*}
    m_t = 173.3~\text{GeV}, \quad m_H = 125~\text{GeV}, \quad m_Z = 91.1876~\text{GeV}, \quad m_W = 79.8244~\text{GeV}.
\end{equation*}
The other fermion masses are taken to be zero. 

The value of $X_{\text{SM}}$ at the beginning of Eq.~\ref{eq:XS_dep} is scaled to the best theoretical prediction for each observable, as documented in the references given in Table~\ref{tab:measurements}.

The fit to data is performed using the open source \HEPfit  package~\cite{deBlas:2019okz, hepfitsite}.
\HEPfit is a general tool designed to combine direct and indirect constraints in the SM \cite{deBlas:2016ojx}, in EFT \cite{deBlas:2018tjm, Durieux:2019rbz} or in particular BSM extensions \cite{Eberhardt:2020dat,Eberhardt:2021ebh}.
Its flexibility allows to easily implement any BSM model or observable. The fit is performed as a Bayesian statistical analysis of the model, in which both theoretical and experimental uncertainties are included. 
\HEPfit includes a Markov Chain Monte Carlo implementation provided by the Bayesian Analysis Toolkit \cite{Caldwell:2008fw} to explore
the parameter space.

\section{Compatibility with the SM predictions}

\begin{figure}[!htbp]
    \centering
    \includegraphics[width=0.45\textwidth]{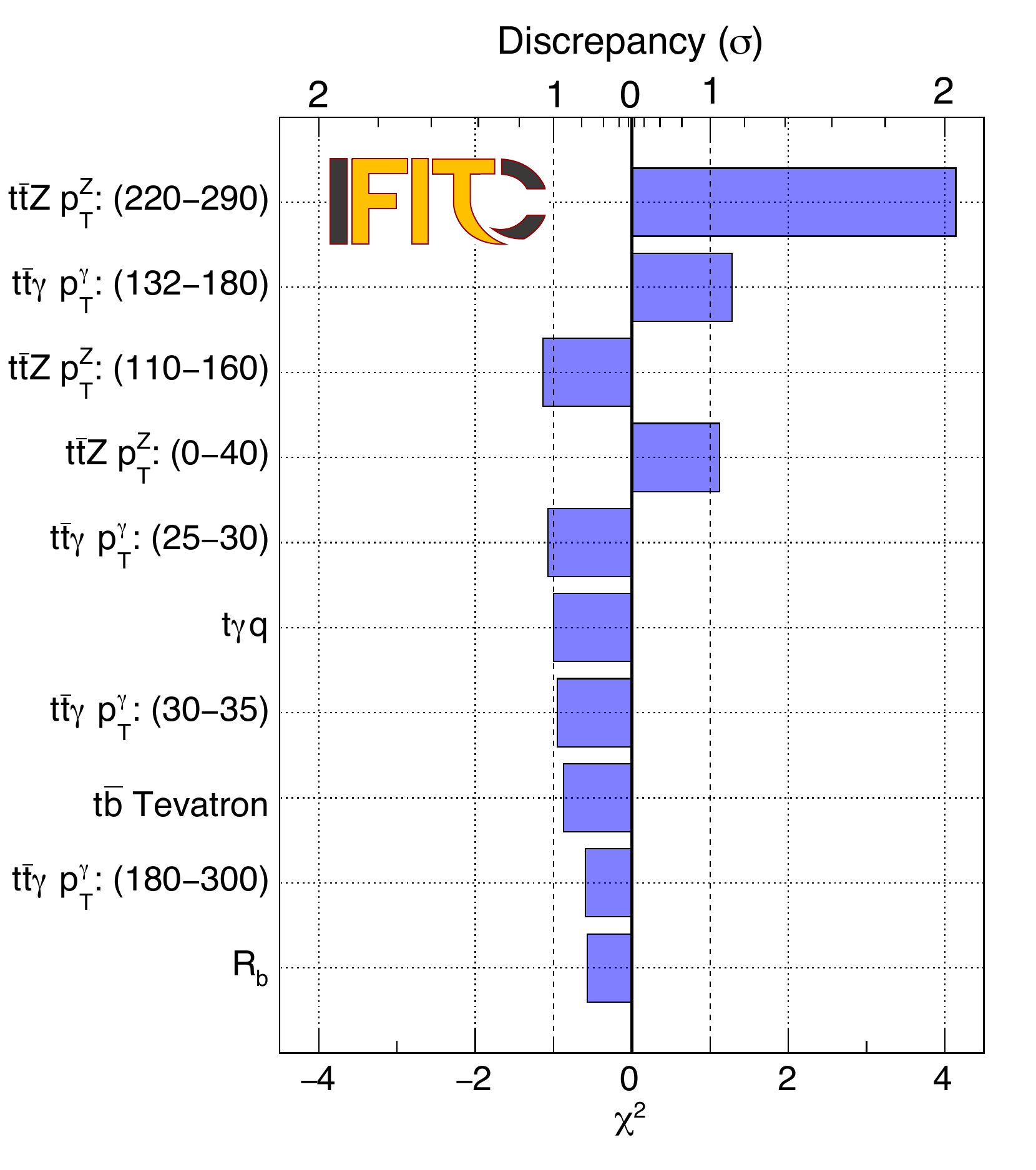}
    \caption{Contributions to the $\chi^2_{\text{SM}}$ of some of the input measurements, together with their discrepancy with respect to the SM predictions. The full fit has $30$ data points and a cumulative $\chi^2_{\text{SM}}\sim21.3$. The negative values indicate that the prediction is lower than the measurement. Both theoretical and experimental uncertainties are considered.}
    \label{fig:SMchi2_plot}
\end{figure}

Prior to the SMEFT fit, the quality of the fit of the selected observables ($n_{\text{obs}}$~=~30) to the SM predictions has been investigated. The correlations among the different data points within a given process, such as the different bins in the $t\bar{t}Z$ and $t\bar{t}\gamma$ differential cross-section measurements, the two $W$~boson helicity fractions ($F_{0}$ and $F_{L}$) and the LEP/SLD observables of the $Zb\bar{b}$ vertex, were published by the experiments and thus are considered in the fit. The obtained chi-square value is $\chi^2_{\text{SM}}/(n_{\text{obs}}-1) = $ 21.3/29, corresponding to a $p$-value of 0.85. Overall, a good agreement
between the SM predictions with the experimental results is seen. However, some observables exhibit some tension with the SM predictions, such as few of the $p_T$ bins of the $t\bar{t}Z$ and $t\bar{t}\gamma$ cross-section measurements. Figure~\ref{fig:SMchi2_plot} shows the $\chi^2_{\text{SM}}$ values for some of the input measurements as well as their discrepancy. 

\section{Interplay between measurements and operator coefficients}
\label{sec:sensitivity}

\begin{figure}[!ht]
    \centering
    \includegraphics[scale=0.4]{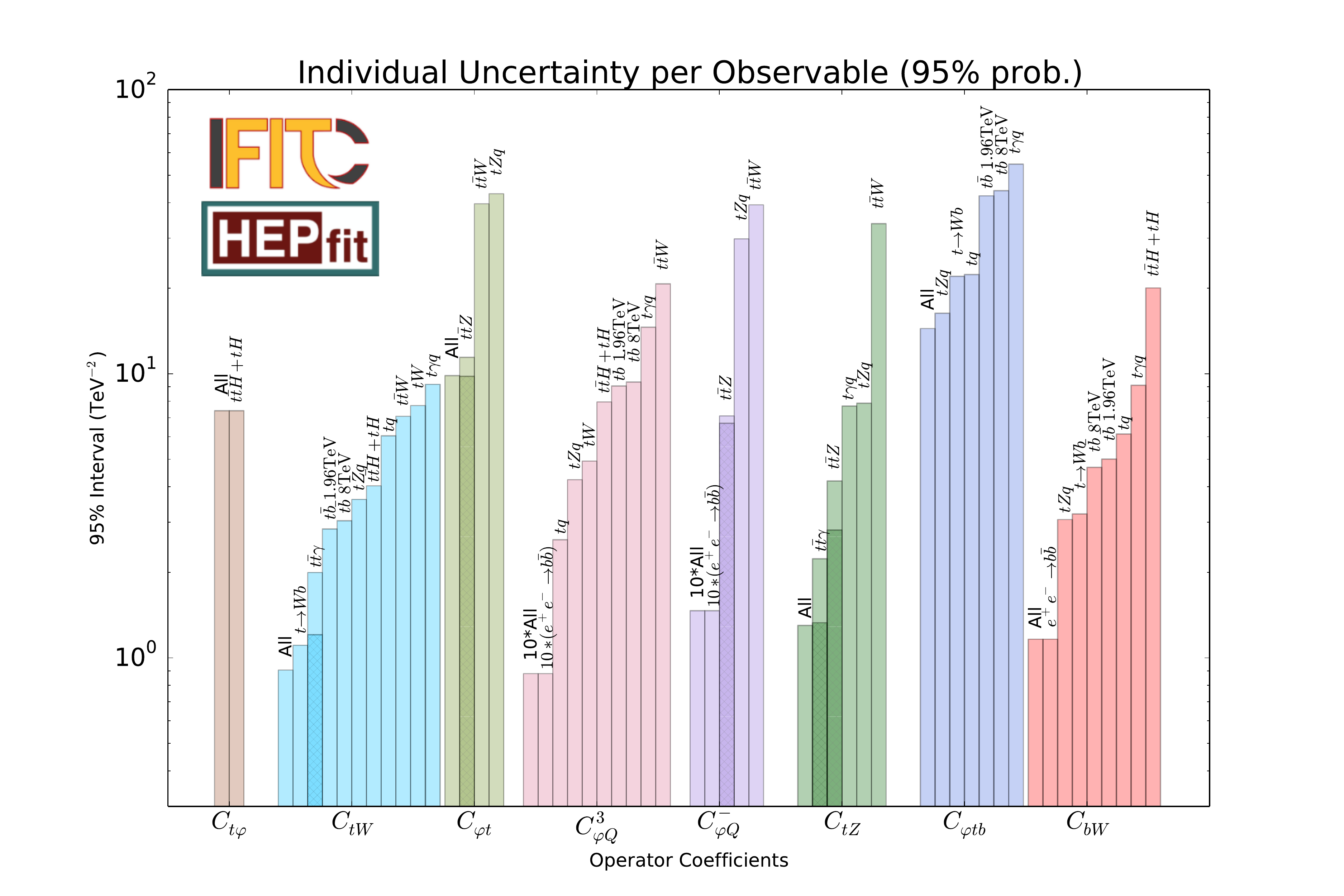}
    \caption{Individual constraints on the eight Wilson coefficient resulting from measurements in different processes. The y-axis corresponds to the full width of the 95\% probability interval for fits including $\Lambda^{-4}$ terms. The observables are set to their SM predictions, maintaining the experimental uncertainty. The dark shading indicates the bound from the $t\bar{t}\gamma$ and $t\bar{t}Z$ differential cross-sections. Besides QCD production, in the case of $t\bar{t}W$ process, the contribution of the electro-weak $t\bar{t}Wq$ production is also included.}
    \label{fig:full_sens}
\end{figure}

Figure~\ref{fig:full_sens} shows the individual 95\% probability bounds on the eight operator coefficients considered in our fit with $\Lambda^{-4}$ terms. The bounds from measurements in different processes, ordered from most to least constraining (going from left to right), are presented for each coefficient. 

The associated production processes of a top quark pair with a $Z$~boson or a photon play an important role in the fit, especially for $C_{\varphi t}$ and $C_{tZ}$ where they provide the leading constraints. For these processes, the bounds obtained from the inclusive cross-section, indicated by the full length of the bar, are compared to the differential measurements, indicated with a darker shading. The bound from the differential measurement is considerably more stringent, especially for $t\bar{t}\gamma$, thanks to the enhanced sensitivity in events where the boson is emitted with a large transverse momentum.

The existing data provide multiple constraints on the different operator coefficients. The one exception is $C_{t\varphi}$, the coefficient that modifies the top quark Yukawa coupling, that is only bounded by the measurement of the $t\bar{t}H$ production rate in our fit. In a fit including more Higgs data, additional constraints arise from the loop diagrams in $gg \rightarrow H$ production and $H\rightarrow \gamma \gamma $ decay\footnote{These two processes give quite competitive individual bounds, but these are not very robust when all other Higgs operators affected by those processes are included in the analysis~\cite{Vryonidou:2018eyv,Durieux:2018ggn,Jung:2020uzh}. Recent global fits of top, Higgs and electro-weak data~\cite{Ellis:2020unq,Ethier:2021bye} confirm this pattern.}.

The $Z$-pole data collected by the LEP/SLD experiments, as included in the electro-weak precision observables~\cite{ALEPH:2005ab}, continue to yield very tight bounds on three operator coefficients: $C_{\varphi Q}^{3}$, $C_{\varphi Q}^{-}$ and $C_{bW}$. The bound on the former two is multiplied by a factor 10 to facilitate the comparison. 

The electro-weak dipole operator $C_{tW}$ is mainly constrained by $W$~boson helicity fractions ($F_0$ and $F_L$) and also by the $t\bar{t}\gamma$ differential cross-sections. The charged current interactions to  right-handed bottom quarks, described by $C_{\varphi tb}$ operator, are difficult to constrain and recent $tZq$ cross-section measurements are as sensitive as the $F_0$ and $F_L$ fractions.\\


\begin{figure}[h!]
    \centering
    \includegraphics[width=0.49\textwidth]{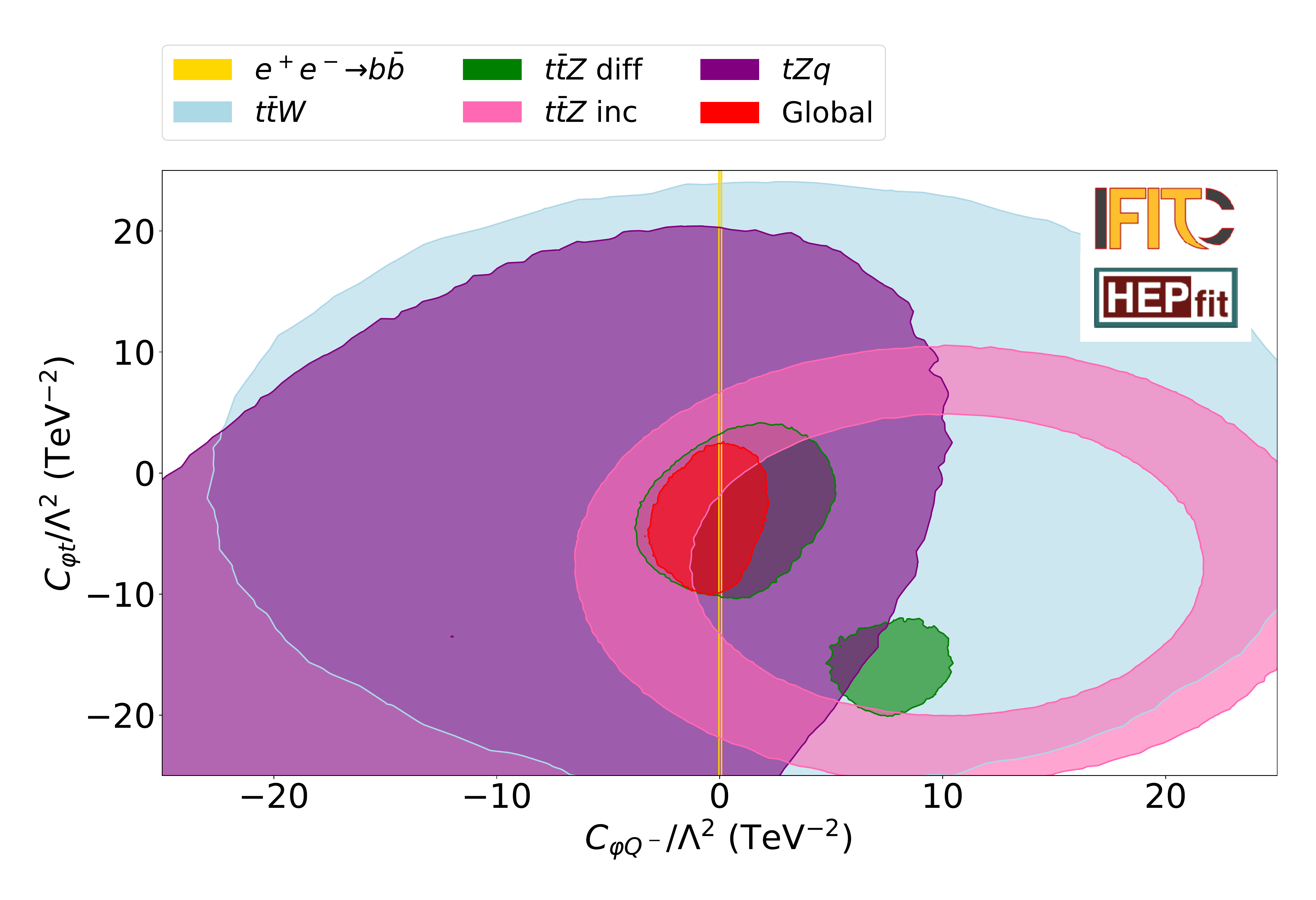} \includegraphics[width=0.49\textwidth]{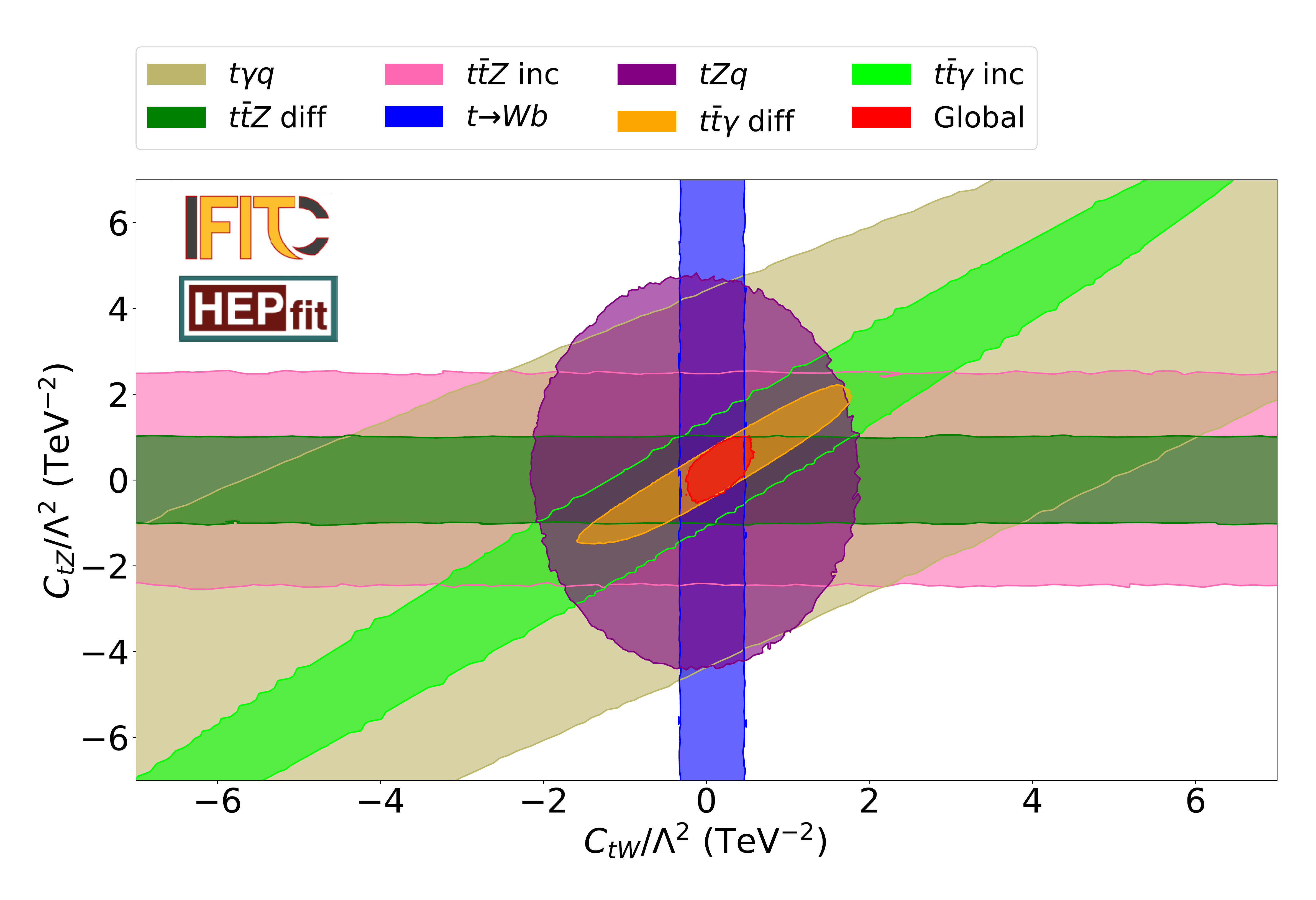}
    \includegraphics[width=0.49\textwidth]{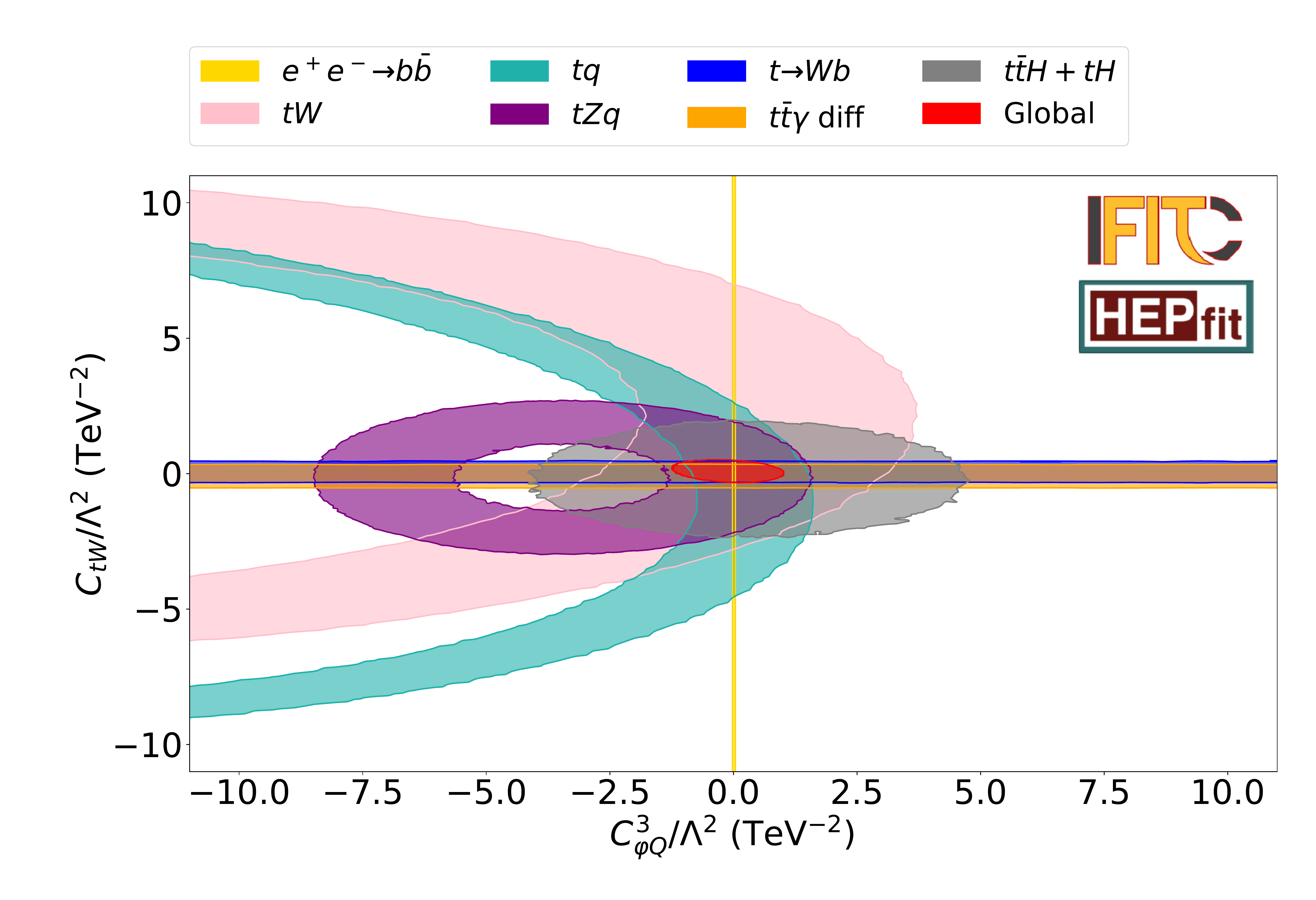}
    \includegraphics[width=0.49\textwidth]{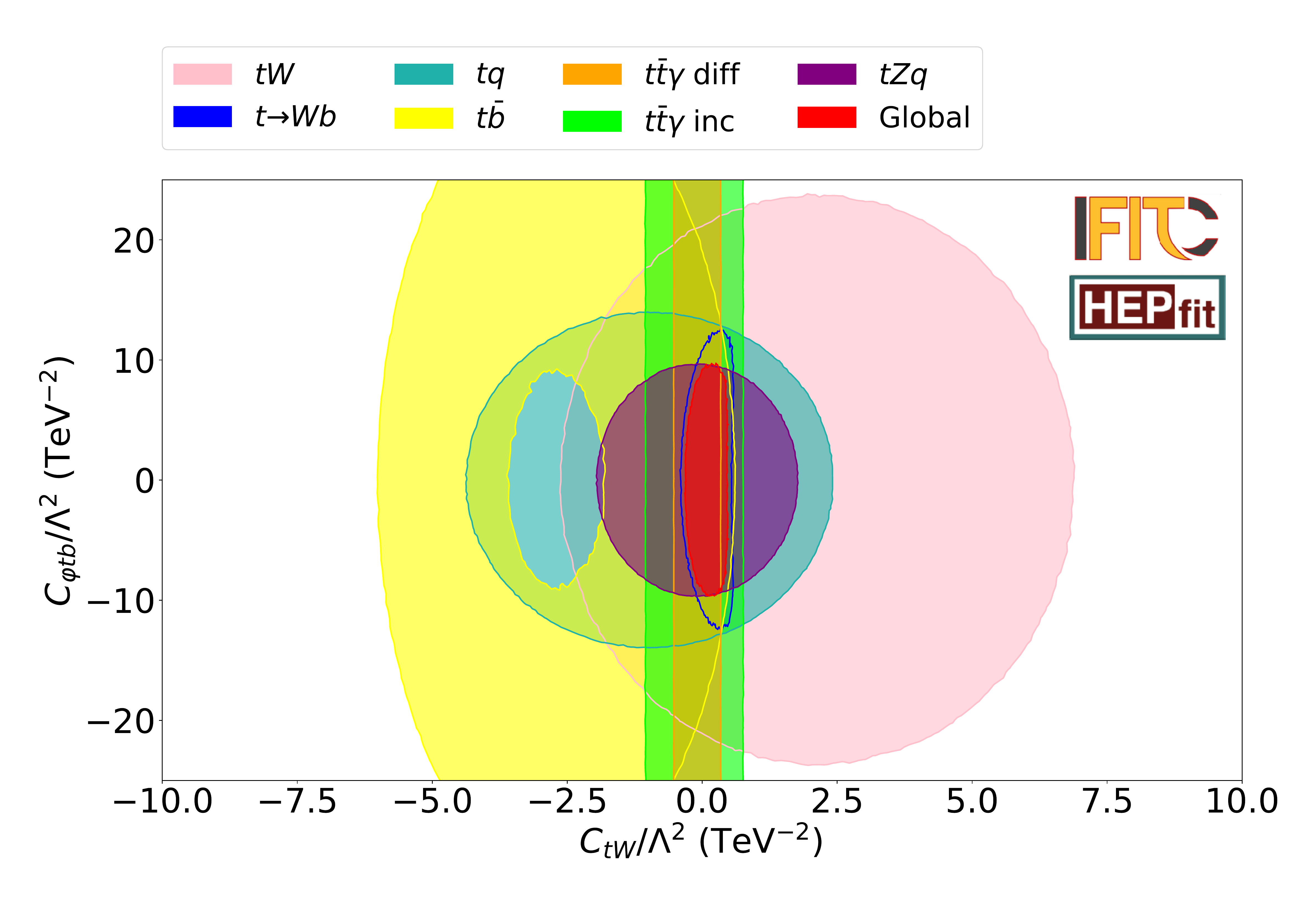}
    \includegraphics[width=0.49\textwidth]{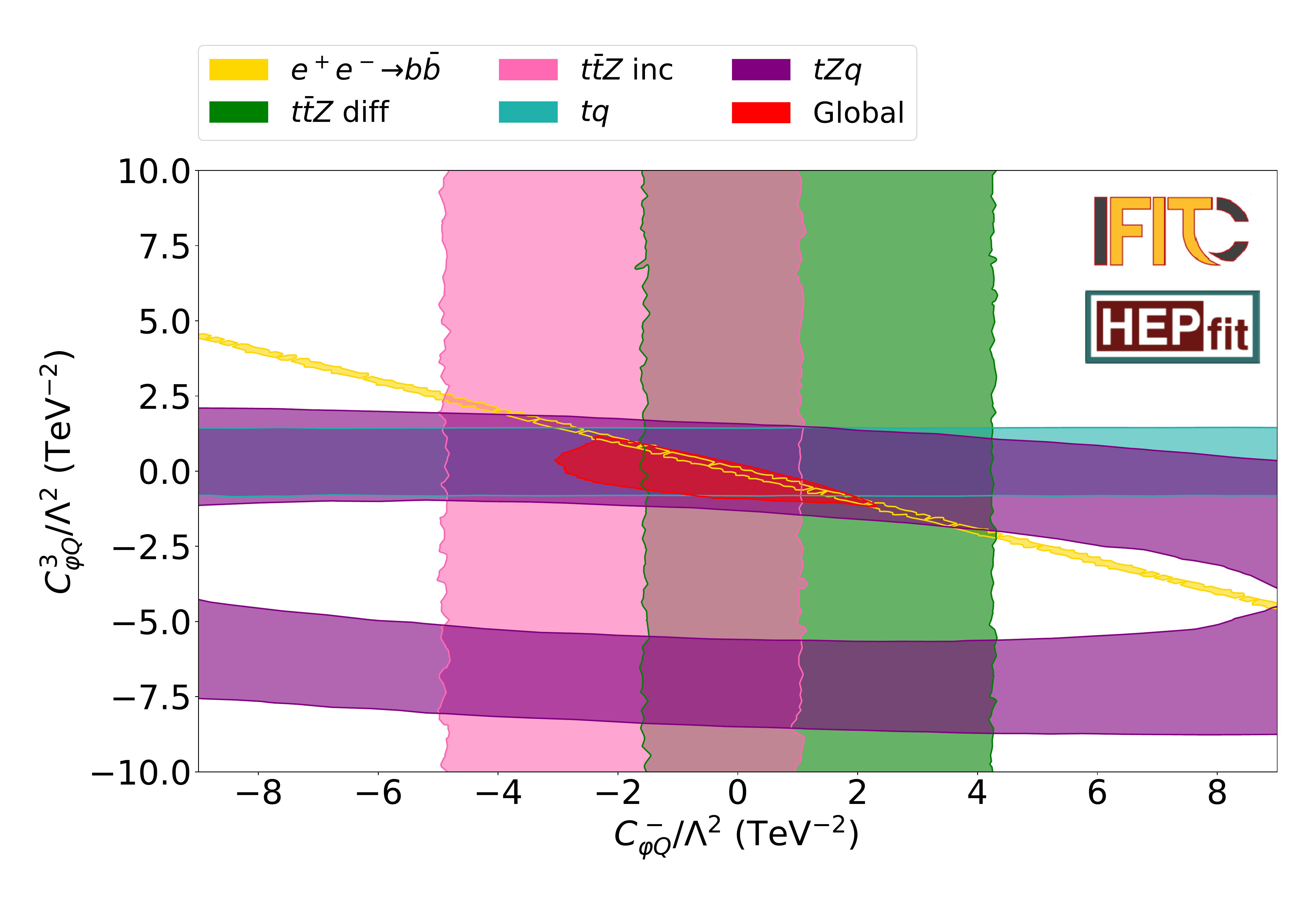}
    \includegraphics[width=0.49\textwidth]{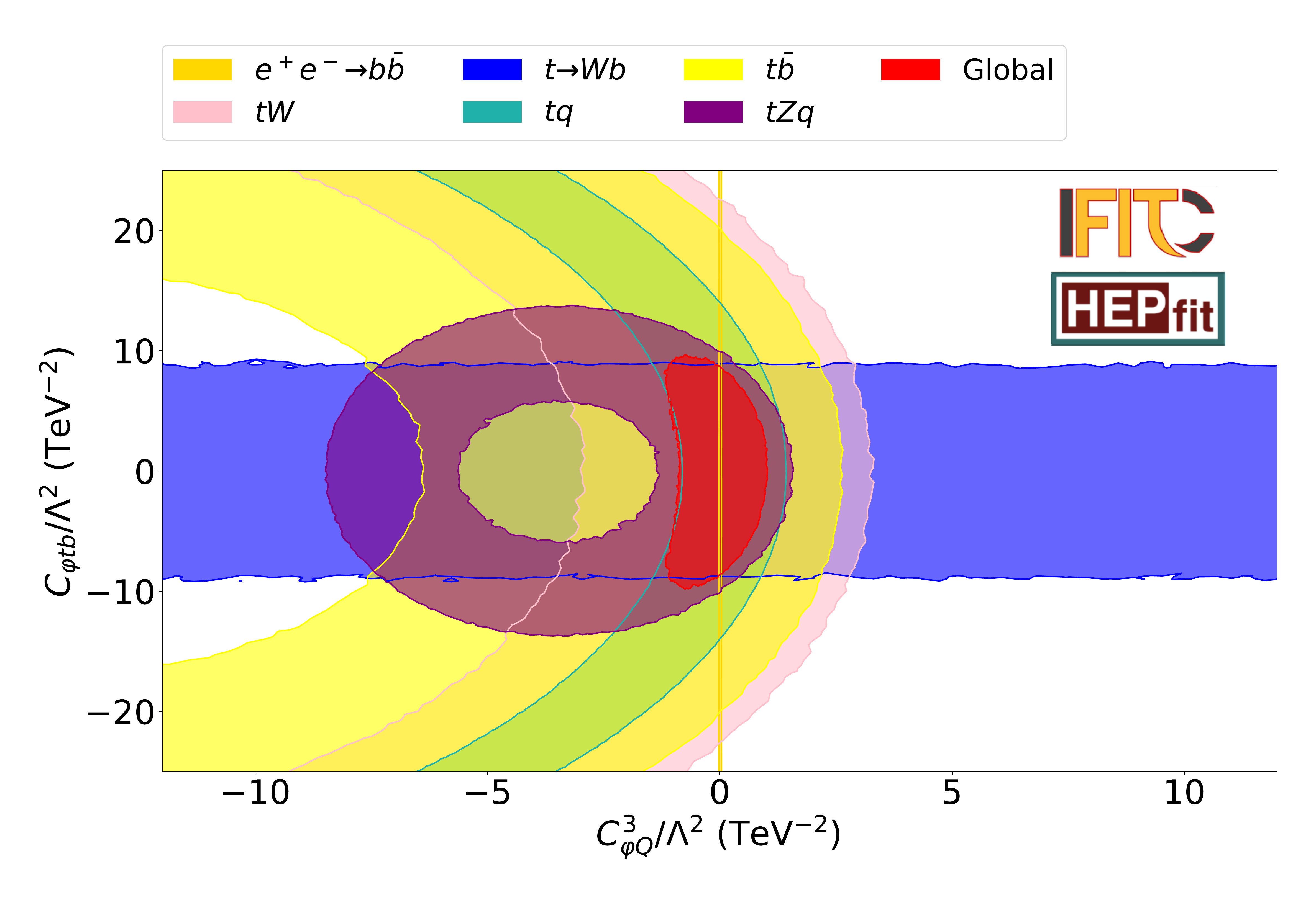}
    \caption{Two-dimensional 95\% probability bounds on pairs of Wilson coefficients, $C_{\varphi Q}^-$ and $C_{\varphi t}$ in the upper leftmost panel, $C_{tW}$ and $C_{tZ}$ in the upper rightmost panel, $C_{\varphi Q}^3$ and $C_{tW}$ in the middle leftmost panel, $C_{tW}$ and $C_{\varphi tb}$ in the middle rightmost panel,   $C_{\varphi Q}^3$ and $C_{\varphi Q}^-$   in the lower leftmost panel, and $C_{\varphi t b}$ and $C_{\varphi Q}^3$ in the lower rightmost panel. Bounds are presented for two-parameter fits to the most constraining measurements. The global fit results, marginalising over all other Wilson coefficients, are also shown (red area). All these fits include $\Lambda^{-4}$ terms. Besides QCD production, in the case of $t\bar{t}W$ process, the contribution of the electro-weak $t\bar{t}Wq$ production is also included.} 
    \label{fig:2D_bounds}
\end{figure}

The power of the measurements to simultaneously constrain multiple Wilson coefficients is illustrated in Figure~\ref{fig:2D_bounds}. The 95\% probability bounds obtained in two-parameter fits to a single measurement are shown as areas of different colours. The red area shows the result of the global fit (ten-operator coefficients), where the result is marginalised over the remaining operator coefficients.

The limits on the left-handed couplings of the $Z$~boson to quarks, rescaled by $C_{\varphi Q}^3$ and $C_{\varphi Q}^-$, are dominated by the LEP and SLD electro-weak precision measurements in $Z\rightarrow b\bar{b}$ decay although they only constrain a linear combination. This linear combination is disentangled once we add other processes like $tZq$, $tq$ or $t\bar{t}Z$.  The strongest limit on $C_{\varphi t}$, right-handed couplings of the quarks to the $Z$~boson, is obtained from the measurements of the associated $t\bar{t}Z$ production cross-section. The $t\bar{t}Z$ cross-section also yields a powerful constraint on $C_{tZ}$ coefficient. The helicity fractions of the $W$~boson in top quark decays provide a tight bound on $C_{tW}$ and $C_{\varphi tb}$, while the $t\bar{t}\gamma$ cross-section provides a complementary constraint on a linear combination of the two dipole operator coefficients.

In both planes in the top, one can see that the differential measurements of the associated $t\bar{t}Z$ and $t\bar{t}\gamma$ production processes improve the bounds very considerably with respect to the inclusive cross-sections. The differential $t\bar{t}Z$ cross-section disentangles $C_{\varphi Q}^-$ and $C_{\varphi t}$, and improves the limit on $C_{tZ}$ by a factor two. The differential $t\bar{t}\gamma$ cross-section restricts the allowed area in the $C_{tW}-C_{tZ}$ plane.

The $tZq$ process and the $W$~boson helicity fractions constrain the $C_{\varphi tb}-C_{tW}$ plane giving complementary limits. A similar behaviour can be seen in the $C_{\varphi tb}-C_{\varphi Q}^3$ plane, where the limits obtained from $tZq$ and $tq$ processes are complementary and once they are combined a much stronger constraint is obtained. 

\section{Bounds of a global EFT fit (to all operator coefficients)}
\label{sec:globalbounds}

\begin{figure}[!ht]
    \centering
    \includegraphics[scale=0.4]{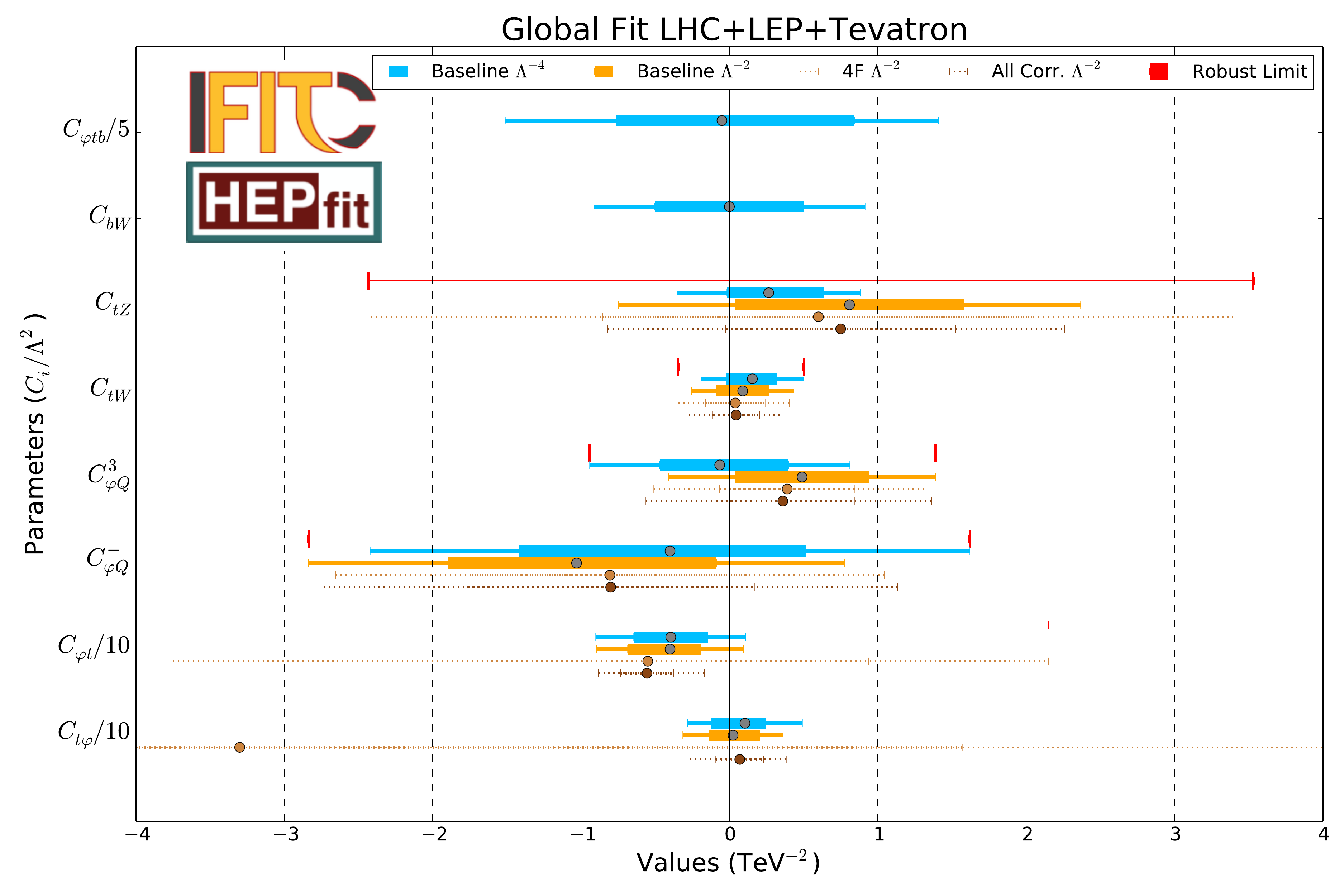}
    \caption{Results of the global fits with $O(\Lambda^{-2})$ (in orange) and $O(\Lambda^{-4})$ (in blue) terms. The two thin lines below each of them correspond to additional fits performed to test the robustness of the results: to account for the effects of the inclusion of further operators and for the correlations between all the different measurements (as described in Section~\ref{sec:robust}). The red markers correspond to the envelope of those additional fits plus another one that accounts for the theoretical uncertainties on the parameterisations.}
    \label{fig:comp}
\end{figure}

The main results of this work are the 68\% and 95\% probability intervals for the Wilson coefficients that modify the top quark electro-weak couplings. Two set of results are given for global fits with $O(\Lambda^{-2})$ and $O(\Lambda^{-4})$ terms, as shown in
Figure~\ref{fig:comp}, in orange and blue respectively. In this figure we also show the result for different stress tests that we have performed to test the robustness of the results. The dotted light brown lines show the result increasing the basis of the linear baseline fit with $C_{tG}$ and seven additional four-fermion operators. The dotted dark brown lines show the result of including correlations between the observables considered in the linear baseline fit. Finally, the red lines show the envelope of all the stress tests that we have performed. The discussion of these stress tests is postponed to the next section. The coefficients $C_{\varphi tb}$ and $C_{bW}$ coefficients can not be constrained in the linear fit since their $X_{i}^{(1)}$ terms in Eq.~\ref{eq:XS_dep} vanish in the limit $m_b \rightarrow 0$.

Overall, the two global fits with $O(\Lambda^{-2})$ and $O(\Lambda^{-4})$ yield comparable results and all of the coefficients are centered at zero within 95\% probability intervals. The difference between the two sets of results is sizeable only for the dipole operator coefficient $C_{tZ}$. This bound is dominated by the $t\bar{t}\gamma$ and $t\bar{t}Z$ differential cross-section measurements, where the term proportional to $\Lambda^{-2}$ is known to be strongly suppressed~\cite{Bylund:2016phk}. A small shift (within 95\% probability intervals) is seen for $C_{\varphi t}$ driven by the sixth $p_T$ bin of the $t\bar{t}Z$ cross-section measurement in which data are below the prediction. In addition, the $C_{\varphi Q}^3$ and $C_{\varphi Q}^-$ coefficients are slightly shifted in opposite directions. The dependence of the fit result on terms proportional to $\Lambda^{-4}$ is significantly reduced compared to previous analyses~\cite{Durieux:2019rbz}. As the LHC measurements become more precise, the bounds remain more and more in the range where the linear terms are dominant and the fit results are valid in full generality.

The obtained limits (95\% probability intervals) for the fitted operator coefficients are presented in Table~\ref{tab:bounds_95_baseline}. For completeness, the limits obtained when fitting only one parameter and fixing the others to zero (individual fits) are also shown in this table. The obtained bounds for $C_{\varphi Q}^3$ and $C_{\varphi Q}^-$ in the individual fits are much better than for the global fits. For the rest of coefficients, the limits do not degrade that much in the global fit.

\begin{table}[!ht]
    \centering
    \resizebox{1.02\columnwidth}{!}{
    \begin{tabular}{c|cc|cc|c}
    $C/\Lambda^2$ & \multicolumn{2}{c|}{Linear (95\% probability)} & \multicolumn{2}{c|}{Lin.+Quad. (95\% probability)}& (95\% probability)\\
     (\text{TeV}$^{-2}$) & Individual     &   Global-Baseline  & Individual     &   Global-Baseline   &  Global-Robust  \\
\hline
    $C_{t \varphi}$ & [-3.17, 3.47]   &  [-3.13, 3.63] & [-3.05, 4.05]   & [-2.82, 4.92] & [-121.82, 62.82]  \\  
    $C_{\varphi Q}^-$ & [-0.038, 0.079]  &    [-2.84, 0.78] & [-0.038, 0.079]  &  [-2.42, 1.62] &  [-2.84, 1.62]   \\
    $C_{\varphi Q}^3$ & [-0.019, 0.040] &  [-0.41, 1.39] & [-0.019, 0.040] & [-0.94, 0.81]  &  [-0.94, 1.39] \\
    $C_{\varphi t}$ & [-6.6, 1.8] &  [-8.96, 0.96]  & [-8.6, 1.5] &  [-9.01, 1.11]    &  [-37.50, 21.50]  \\
      $C_{tW}$ & [-0.30, 0.38]   & [-0.26, 0.44]  & [-0.28, 0.32]  &  [-0.19, 0.50] & [-0.35, 0.50] \\
    $C_{tZ}$  & [-0.82, 2.21]  &   [-0.75, 2.37] & [-0.39, 0.57]  &   [-0.35, 0.88]  &   [-2.43, 3.53]  \\
       $C_{\varphi tb}$ & -- & -- & [-6.61, 6.71] &  [-7.55, 7.05] &   --  \\
    $C_{bW}$ & -- & --  & [-0.47, 0.47] &  [-0.91, 0.91]   &  -- \\
    \end{tabular}
    }
      \caption{ Allowed ranges of the Wilson coefficients with a probability of 95\% expressed in TeV$^{-2}$ including only linear terms or linear and quadratic terms. We show, from left to right, the results of five fits: individual with linear terms, global baseline with linear terms, individual with linear and quadratic terms, global baseline with linear and quadratic terms and global robust limits. The robust result accounts for the effects of the correlations between the observables, the inclusion of further operators and the theoretical uncertainties on the parameterisations.} 
    \label{tab:bounds_95_baseline}
\end{table}

The EFT fit with terms $O(\Lambda^{-4})$, with ten floating Wilson coefficients, returns a log-likelihood value $\log{L_{\text{EFT}}}= -10.1$ (see Table~\ref{tab:statCOMP}) and the additional degrees of freedom do not improve the already excellent agreement of the fit. The observables that contribute most to the log-likelihood are the differential cross-section measurements of $t\bar{t}Z$ ($\Delta \log{L}= 5.8$) and $t\bar{t}\gamma$ ($\Delta \log{L} = 2.7$) processes as a function of the boson transverse momentum, with 7 and 11 bins respectively. Similar values are obtained for the fit with $O(\Lambda^{-2})$ terms.

\begin{table}[!ht]
    \centering
    \begin{tabular}{|c|c|c|c|c|c|}
    \hline
 Fit &    Correlation & $\log L$ & $\chi^2/\text{d.o.f}$ & $p$-value \\ 
      \hline
 SM   &  All published &  -10.6 & 21.3/29 & 0.85\\
 SM   &  + our ansatz   &  -11.6 & 23.2/29 & 0.77\\ \hline
 EFT Quad. &  All published &  -10.1  & 20.1/19 & 0.39\\
 EFT Quad. &  + our ansatz   & -10.2 & 20.4/19 & 0.37 \\
        \hline
EFT Lin. &  All published & -10.7 & 21.5/22 & 0.49\\
 EFT Lin. &  + our ansatz   & -11.0  & 22.1/22 & 0.45 \\
        \hline
    \end{tabular}
    \caption{Values of $\log L, \chi^2/\text{d.o.f}$ and $p$-values for different assumptions on the correlations among the measurements in the SM and EFT fits. ``All published'' include the theoretical correlations and the published experimental correlations of the $t\bar{t}Z$, $t\bar{t}\gamma$, $F_0-F_L$ and LEP/SLD measurements. The number of degrees of freedom (d.o.f.) is the number of observables ($n_{\text{obs}}$) minus the number of fitted parameters and minus one. We assume Gaussian errors to extract the $\chi^2$ from $\log L$.}
    \label{tab:statCOMP}
    \end{table}

A convenient metric to quantify the strength of the constraints in the $n$-dimensional parameter space of effective-operator coefficients is the so-called global determinant parameter (GDP) defined as the $2n$ root of the Gaussian covariance matrix determinant~\cite{Durieux:2017rsg}:  GDP~=~$\sqrt[2n]{\det V}$. This figure of merit measures the hypervolume of the allowed parameter space, taking into account the correlations among coefficients. As it assumes that the errors are Gaussian, we cannot use this determinant in the fit including $O(\Lambda^{-4})$ terms but it is still interesting to show the results for the fit including only $O(\Lambda^{-2})$ terms. For the partial Run 2 result that we reported in 2019~\cite{Durieux:2019rbz} we obtain a GDP score of 1.37. The new Run~2 data included in the current fit improves the GDP to 0.58 with inclusive measurements only, and to 0.49 when the differential measurements are used. The bounds on the Wilson coefficients are improved, on average, to 70\% thanks to the addition of the new LHC Run~2 measurements.

\section{Robust bounds}
\label{sec:robust}

In this section, a number of effects have been investigated to test the validity of the obtained limits. We assess the quantitative impact
of the inclusion of further degrees of freedom of the SMEFT, of correlations among the different observables, and of missing higher-orders (uncertainty on the scale choice) for the parameterisation between the observables and the Wilson coefficients.

\subsection{Theory uncertainties on the parameterisation}
\label{sec:theoryuncparam}
Theoretical uncertainties due to missing higher orders in $\alpha_s$ in the predictions for the terms $X_i^{(1)}/X_{SM}$ and
$X_{ij}^{(2)}/X_{SM}$ in Eq.~\ref{eq:XS_dep}, that parameterise the dependence of the observables on the Wilson coefficients, have also been considered. However, this effect does not have an important impact on the fit (maximum of 5$\%$) given that most of the predictions are extracted at NLO.

\subsection{Limitations of the basis}
\label{sec:limitations}

The set of operators included in the fit ignores the impact of the two-fermion operator $C_{tG}$ and of the four-fermion operators with two light quarks and two heavy quarks. In global fits, these are constrained by the precise $t\bar{t}$ cross-section measurements at the Tevatron and LHC, but may still have an impact on the $t\bar{t}X$ production cross-sections. The sensitivity to these operator coefficients is assessed by extending the operator basis with $C_{tG}$ and the seven four-fermion operators that affect the $t\bar{t}X$ processes included in the analysis via $O(\Lambda^{-2})$ terms at LO QCD \cite{Brivio:2019ius}. The Wilson coefficients of these four-fermion operators are
$$
C_{tu}^8,\,C_{td}^8,\,C_{Q q}^{1,\,8},\,C_{Q q}^{3,\,8},\,C_{Qu}^8,\,C_{Qd}^8,\,C_{tq}^8,
$$
which are defined with the same notation of Ref.~\cite{Brivio:2019ius}.

To provide a constraint on these additional degrees of freedom, the inclusive $p\bar{p} \rightarrow t\bar{t}$ cross-section measurement from the Tevatron experiments~\cite{CDF:2013hmv} and the $pp\rightarrow t\bar{t}$ cross-section at 13~TeV~\cite{ATLAS:2019hau} are included in this extended fit. In order to perform a competitive global fit on these additional degrees of freedom we should include more observables, like differential measurements of $pp\rightarrow t\bar{t}$, the $t\bar{t}$ charge asymmetry~\cite{Rosello:2015sck} or the top energy asymmetry~\cite{Basan:2020btr}. The purpose of this analysis is just to see the degradation of our limits due to extending the basis, not to find a competitive constraint on $C_{tG}$ and/or the four-fermion operators. 

Looking at the dotted light brown lines of Figure~\ref{fig:comp}, we clearly see how extending the basis has a slight effect on $C_{tW}$, $C_{\varphi Q}^3$ and $C_{\varphi Q}^-$. The limits on $C_{tZ}$ are degraded by a factor of two but they are still competitive. For $C_{\varphi t}$ and $C_{t \varphi }$ the limits are totally ruined. Indeed, these three operators ($C_{tZ}$, $C_{\varphi t}$ and $C_{t \varphi }$) that are mostly affected by extending our basis
 are mainly constrained by $t\bar{t}X$ cross-section measurements in the baseline fits, as can be seen in Figure~\ref{fig:full_sens}. 
 Precisely, these cross-sections are the ones affected by the presence of $C_{tG}$ and the four-fermion operators. 

Operators with two heavy quarks and two charged leptons (type $t\bar{t}l^+l^-$) are also ignored in our fit. Their impact on the operators that are constrained by the measurements in the $t\bar{t}Z$ process must be assessed carefully in future work~\cite{Sirunyan:2020tqm}.

\subsection{Correlations between measurements}
\label{sec:correlations}

The selection of measurements included in our fit aims to minimise the statistical correlation among them. However, the correlation of experimental and theoretical systematic uncertainties among the different processes could be sizeable.
Since such correlations have not been provided by the experiments, we vary them from zero to the ansatz discussed here.

For the experimental correlations, we estimate an ansatz as follows. For each pair of measurements we break down the uncertainties reported in the experimental publications into: i) statistical, ii) mildly correlated, and iii) highly correlated systematic uncertainties. The combined correlation between two observables $i$ and $j$ is then given by: 

\begin{equation}
    \bar{\rho}_{ij}^{\alpha} = \frac{\sum_{\alpha} \rho_{ij} u_i^{\alpha} u_j^{\alpha}}{ \sqrt{\left(\sum_{\alpha} u_i^{\alpha} u_i^{\alpha} \right) \left(\sum_{\alpha} u_j^{\alpha} u_j^{\alpha} \right)}} \, ,
\end{equation}
where $u_i^{\alpha}$ is the uncertainty of the observable $i$, $\alpha$ runs over the three categories of uncertainties and $\rho_{ij}$ is an ansatz which varies from 0 to 0.5 (depending on the source of the uncertainty).

In addition, theoretical uncertainties on the predictions can be correlated, due to the PDFs and the similarity of certain production processes. For the observables measured at the LHC, the ansatz for this correlation is: 100\% between the different bins of $t\bar{t}Z$ and $t\bar{t}\gamma$ differential cross-sections, $-100$\% between the two $W$~helicity fractions $F_{0,L}$, and either 20\% or 50\% between the other 13~TeV measurements and some 13~TeV - 8~TeV and 8~TeV - 8~TeV pairs. The correlations of a given observable with $F_L$ have opposite sign compared to the correlations of such observable with $F_0$.

The ansatz for the correlations is a rough estimate, but allows us to investigate the robustness of the result. The most significant effects are the shifts observed for $C_{\varphi t}$, $C_{\varphi Q}^-$ and $C_{\varphi Q}^3$ coefficients. The shifts are small compared to the width of the 95\% probability interval, which are $\cal{O}$(1 \TeV$^{-2}$) for the $C_{\varphi Q}^-$ and $C_{\varphi Q}^3$ coefficients and $\cal{O}$(10 \TeV$^{-2}$) for $C_{\varphi t}$. The shifts on $C_{\varphi t}$ increase the tension with the SM. The log-likelihood of the fit in Table~\ref{tab:statCOMP} goes to larger negative values as the degree of correlation among the measurement increases, but this corresponds to a very mild decrease in the $p$-values.

\subsection{Discussion}

We have attempted to perform a robust analysis and have presented several stress tests to assess the impact of the limitations of EFT analyses.
The ``robust limits", obtained as the envelope of all these fits, are indicated with red vertical lines in Figure~\ref{fig:comp} and in the right column of Table~\ref{tab:bounds_95_baseline}.

In the baseline fits we have included all published correlations between measurements, taking advantage in particular of LHC combinations whenever available.
To avoid strong - and unknown - correlations between measurements, we have included only the most precise measurement of each observable, discarding less precise measurements by other experiments, and at other centre-of-mass energies. We have also checked that, whenever available, including measurements at two center-of-mass energies, has a negligible impact in our limits. This reinforces our strategy of including them only at one center-of-mass energy.
To estimate the residual impact of correlations, we include a complete covariance matrix with plausible ansatz values. The fit is found to be relatively robust to the assumed correlations, with minor changes in the results for three operator coefficients. Compared to Ref.~\cite{Bissmann:2019gfc}, correlations seem relatively harmless. We consider that this is due to our choice to select a single result among redundant measurements. Eventually, with the legacy combinations by the ATLAS and CMS collaborations, this unsatisfactory reduction of the data set will no longer be necessary.

\section{Conclusions}

The most precise measurements of top quark associated production with bosons from ATLAS and CMS experiments at LHC, the latest LHC combined results of $W$~boson helicity fractions and single top quark production, as well as a few legacy LEP/SLD and Tevatron measurements have been considered to perform a global analysis of the top quark electro-weak sector in the SMEFT. 

We find that the Standard Model offers an adequate description of the 30 observables included in the fit. The total $\chi^2/\text{d.o.f.}$ is 21.3/29, corresponding to a $p$-value greater than 80\%. The 95\% probability intervals of the SMEFT fit include the SM prediction ($C=0$) for the eight Wilson coefficients that modify the top quark electro-weak couplings. The global 95\% intervals vary from $\pm 0.35$ to $\pm 8$ \TeV$^{-2}$.

The results of our fit improve significantly with respect to a previous analysis of partial Run~2 data~\cite{Durieux:2019rbz}. This demonstrates the value of recent LHC results based on analyses of the complete Run~2 data set. In particular, the differential cross-section measurements for the $t\bar{t}Z$ and $t\bar{t}\gamma$ processes represent a significant step forward.

Following LHC Top Working Group recommendations~\cite{AguilarSaavedra:2018nen}, we report the results of two EFT fits, with parameterisations including only terms proportional to $\Lambda^{-2}$ or with both $\cal{O}$$(\Lambda^{-2})$ and $\cal{O}$$(\Lambda^{-4})$ terms. The largest difference between the two fits is observed for the coefficient $C_{tZ}$, where the bound still relies strongly on quadratic $\cal{O}$$(\Lambda^{-4})$ terms (as signalled already by Ref.~\cite{Bylund:2016phk}, the $\Lambda^{-2}$ term is suppressed in $t\bar{t}X$ production). 

We have made an important effort to study the resilience of the bounds against uncertainties inherent in the EFT framework. We present robust bounds that take into account the uncertainties due to missing higher orders in $\alpha_s$, potential correlations among the different observables and missing degrees of freedom. The analysis is found to be quite robust against the first two effects but only robust for some operators against the inclusion of other degrees of freedom. The most sensible observables to the extension of the basis are $C_{t\varphi}$,  $C_{\varphi t}$ and $C_{tZ}$, although competitive limits can still be found for the latter coefficient. This result is in agreement with what is observed in global SMEFT fits~\cite{Ellis:2020unq, Ethier:2021bye}.

We take into account the effect of published correlations between measurements. Large and unknown correlations are avoided by selecting a single measurement for each observable. For pairs of measurements where no public information is available, we estimate a plausible ansatz for the correlation. The resulting shifts of the central value are small, which shows that the EFT analysis is reasonably robust. Of course, a rigorous treatment of carefully estimated correlation is a necessary ingredient for a SMEFT analysis of LHC legacy measurements that aims to reach the ultimate precision.

The bounds reported in this paper agree well with the results of more comprehensive analyses that include data on top quark and Higgs boson production and electro-weak precision measurements~\cite{Ellis:2020unq, Ethier:2021bye}, once the differences in the operator basis and treatment of $\Lambda^{-4}$ terms are accounted for. The main exception is the bound on $C_{tZ}$ that is significantly stronger in our analysis. This difference can be traced back to the omission of the $t\bar{t}\gamma$ measurement in Ref.~\cite{Ethier:2021bye} and in the initial result of Ref.~\cite{Ellis:2020unq}. On the other hand, Refs.~\cite{Ellis:2020unq, Ethier:2021bye} report tighter individual bounds on the coefficients $C_{\varphi t}$ and $C_{t\varphi}$ from the Higgs and electro-weak data. This effect remains limited to individual bounds: the global bounds on the top electro-weak couplings remain clearly dominated by top physics data. The interplay between Higgs, electro-weak and top physics (and flavour~\cite{Bruggisser:2021duo}) definitely merits further exploration.

\section*{Acknowledgements}
This project builds heavily on previous work in collaboration with Gauthier Durieux (CERN) and Cen Zhang (IHEP). We acknowledge their help and support in setting up the EFT fit. Cen Zhang passed away unexpectedly in June 2021 and will be missed dearly. We would also like to thank Antonio Pich for his useful comments on the manuscript. This work is supported by the projects: FPA2017-84445-P, PGC2018-094856-B-100 and RTI2018-100863-B-I00 (Spanish Ministry MICINN and FEDER); PROMETEO-2018/060 and SEJI-2020/037 (Generalitat Valenciana); LCF/BQ/PI19/11690014 (La Caixa Banking Foundation); and the iLINK grant LINKB20065 (CSIC). The work of V.M. is supported by the FPU doctoral contract FPU16/0191, funded by the Spanish Ministry MICIU.

\newpage
\section*{Appendix: Correlation Matrices}





\begin{figure}[!htbp] 
    \centering
    \includegraphics[width=\textwidth]{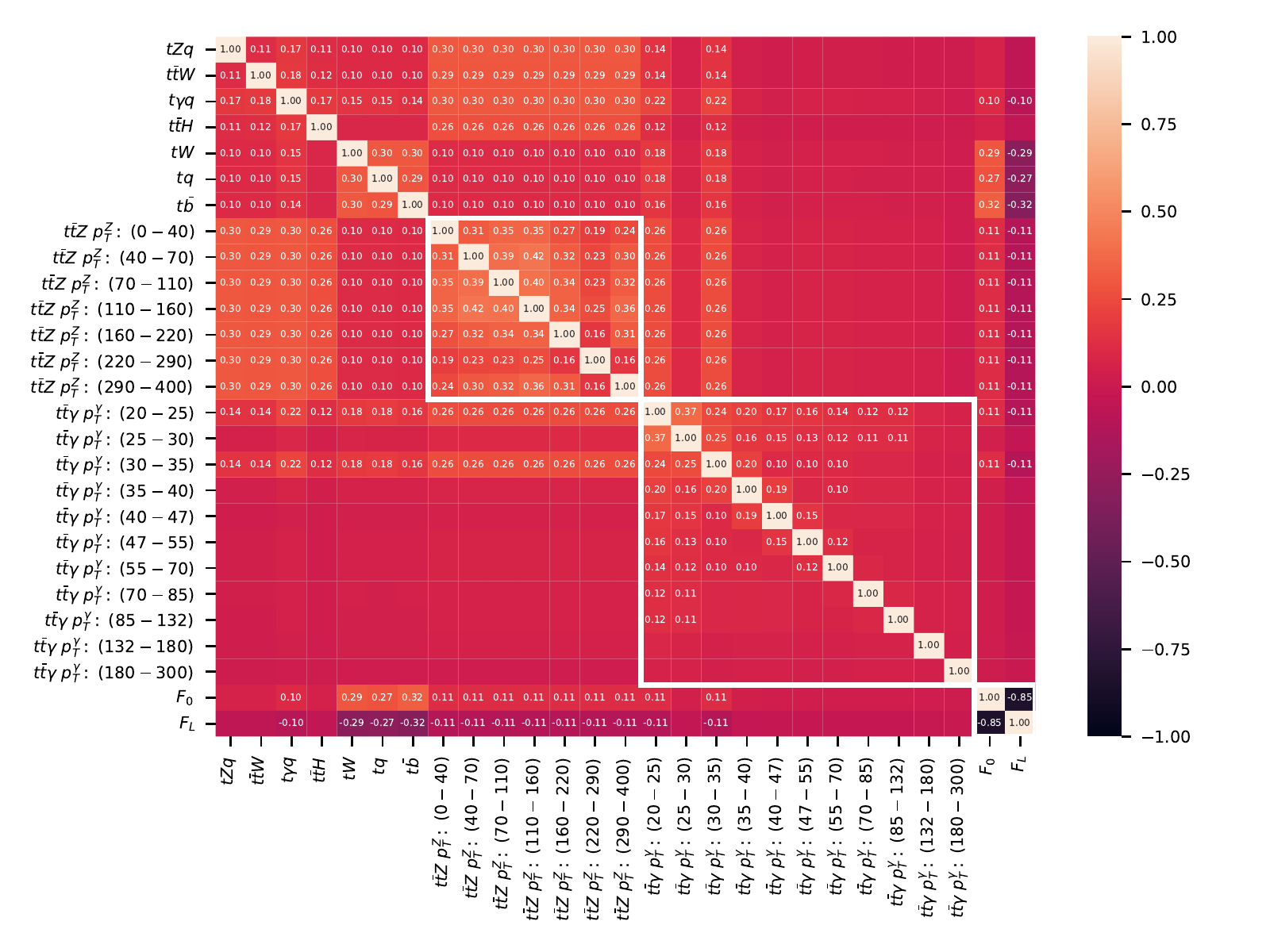}
    \caption{Experimental correlation matrix used. The boxes in white correspond to the correlations published by the experiments for $t\bar{t}Z$, $t\bar{t}\gamma$ and $W$ boson helicity fractions~\cite{ATLAS:2020cxf,Aad:2020axn,Aad:2020jvx}. The rest of the entries correspond to our ansatz, as described in Section~\ref{sec:correlations}. Cells are filled if the correlation is higher than $10\%$ in absolute value.}
    \label{fig:ansatz_exp}
\end{figure}

\begin{figure}[!htbp] 
    \centering
    \includegraphics[width=\textwidth]{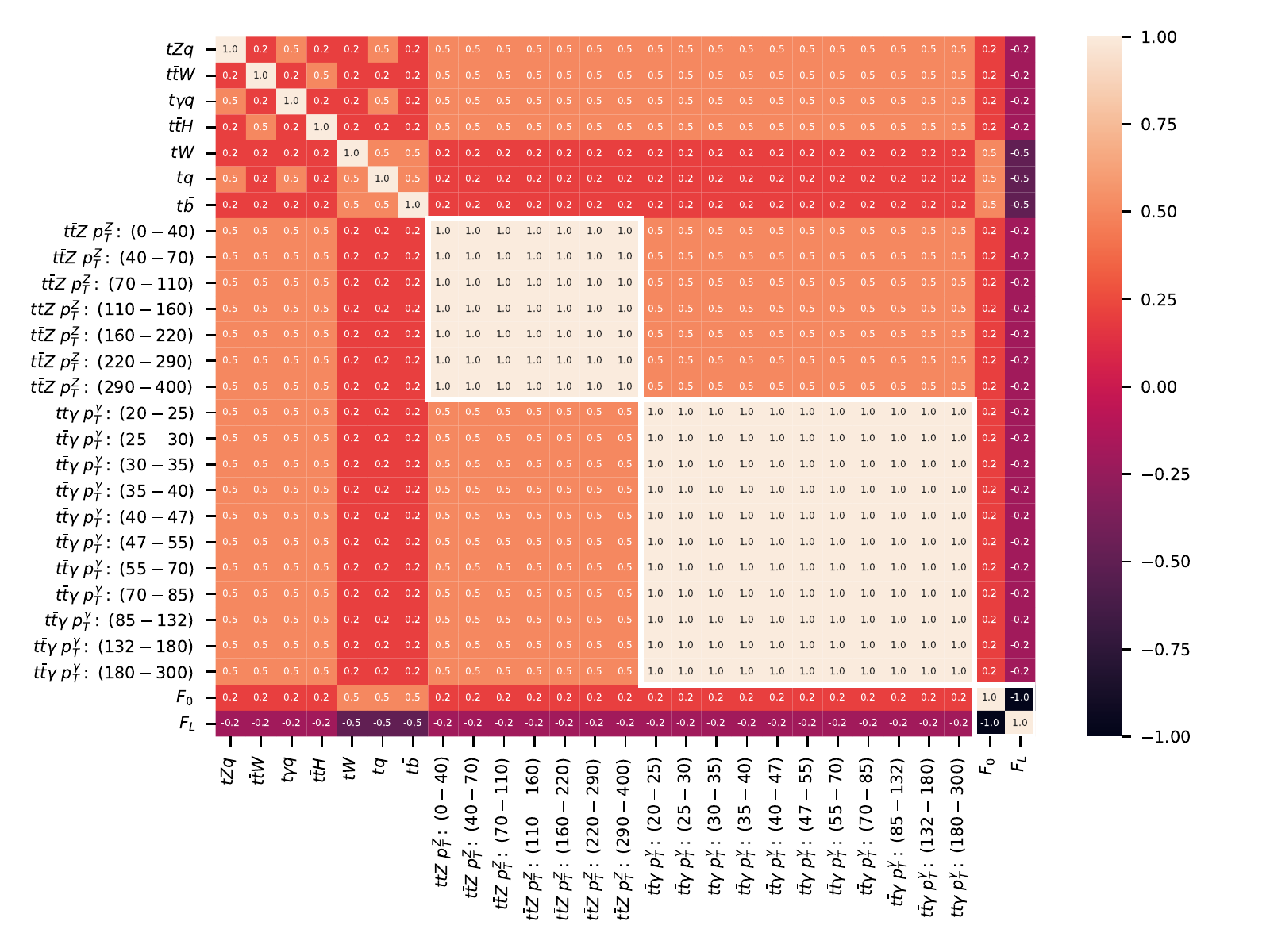}
    \caption{Theoretical correlation matrix used. The boxes in white correspond to the correlations between the differential bins in $t\bar{t}Z$ and $t\bar{t}\gamma$ ($W$ boson helicity fractions) and a 100\% ($-100$\%) correlation is assumed among them. The rest of the entries correspond to our ansatz, as described in Section~\ref{sec:correlations}. Cells are filled if the correlation is higher than $10\%$ in absolute value.}
    \label{fig:ansatz_teo}
\end{figure}

\begin{figure}[!htbp] 
    \centering
    \includegraphics[width=0.7\textwidth]{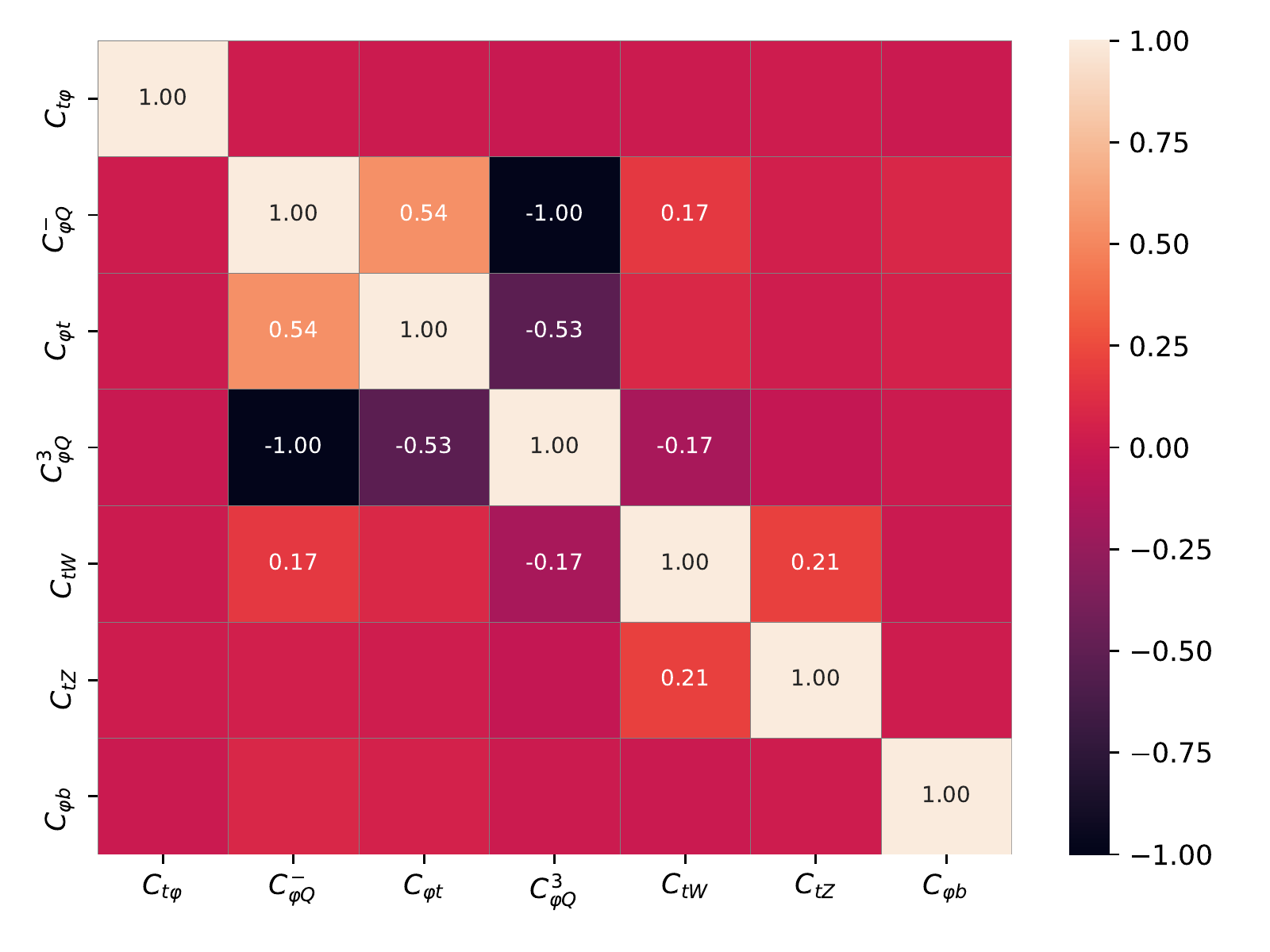}
    \caption{Correlation matrix between the different EFT operators obtained in the baseline linear $(\Lambda^{-2})$ fit. Cells are filled if the correlation is higher than $10\%$ in absolute value. The operator $O_{\varphi b}$, that modifies only the bottom quark electro-weak couplings, is taken into account in the fit but limits on its coefficients are not reported since the obtained values are not competitive using only the observables considered in our fit.}
    \label{fig:out_base_lin}
\end{figure}

\begin{figure}[!htbp]
    \centering
    \includegraphics[width=0.68\textwidth]{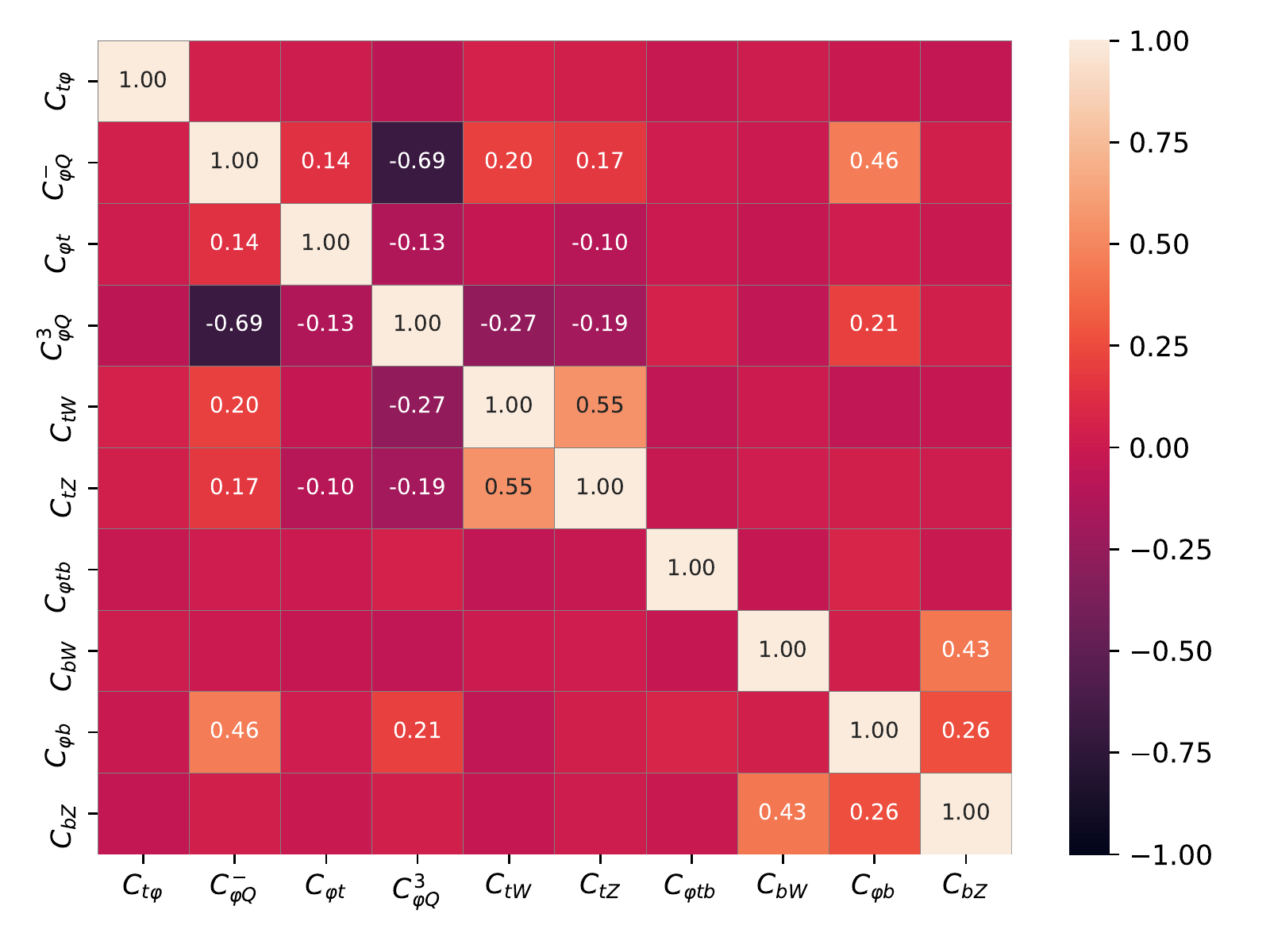}
    \caption{Correlation matrix between the different EFT operators obtained in the baseline quadratic $(\Lambda^{-4})$ fit. Cells are filled if the correlation is higher than $10\%$ in absolute value. The operators that modify only the bottom quark electro-weak couplings, $O_{\varphi d}$ and $O_{dZ}$, are taken into account in the fit but limits on their coefficients are not reported since the obtained values are not competitive using only the observables considered in our fit.}
    \label{fig:out_base_quad}
\end{figure}

\begin{figure}[!htbp] 
    \centering
    \includegraphics[width=0.68\textwidth]{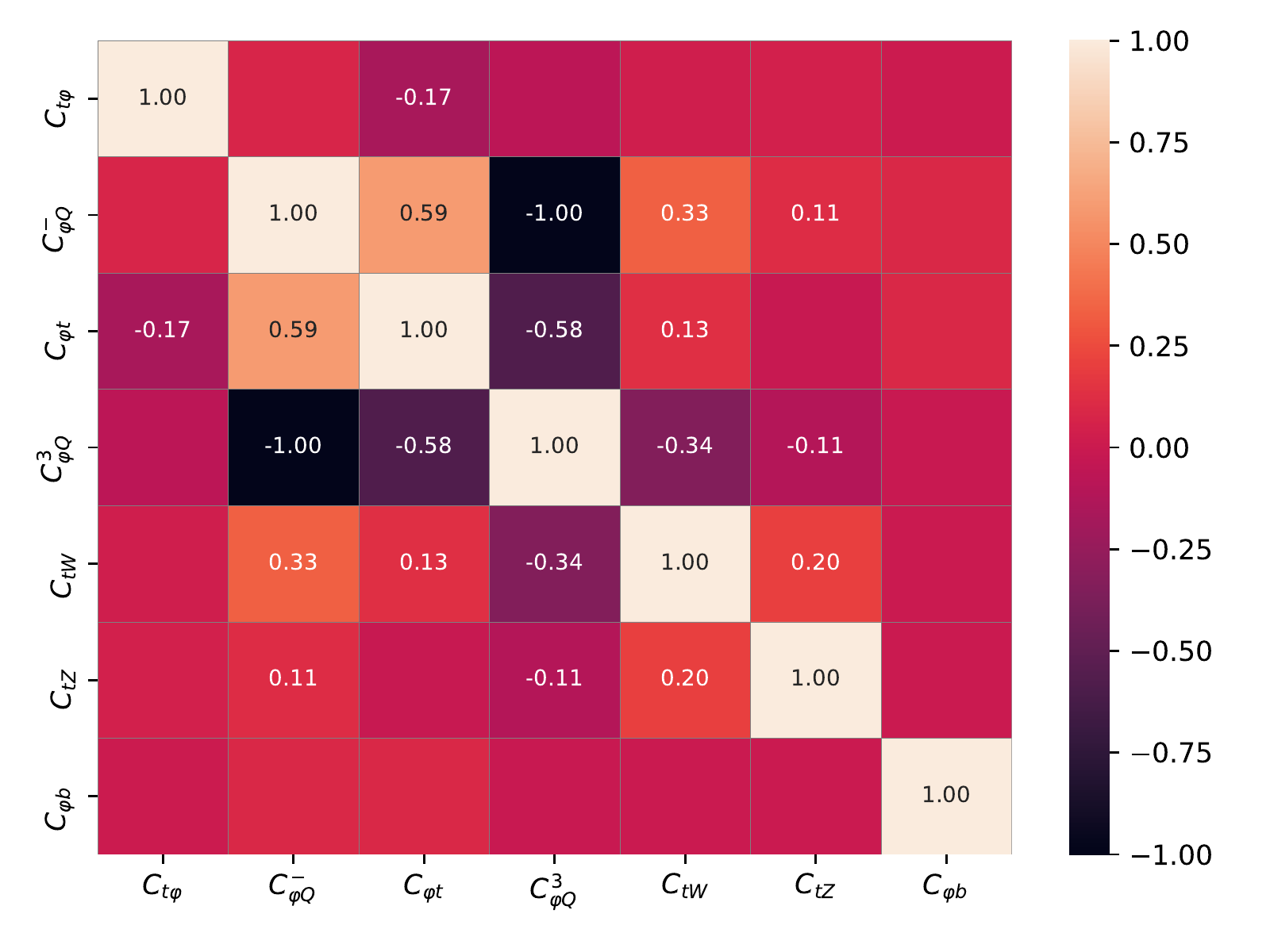}
    \caption{Correlation matrix between the different EFT operators obtained in the linear $(\Lambda^{-2})$ ``All correlations" robustness test fit described in Section~\ref{sec:correlations}. Cells are filled if the correlation is higher than $10\%$ in absolute value. The operator $O_{\varphi b}$, that modifies only the bottom quark electro-weak couplings, is taken into account in the fit but limits on its coefficients are not reported since the obtained values are not competitive using only the observables considered in our fit.}
    \label{fig:out_base_lin_all}
\end{figure}

\begin{figure}[!htbp] 
    \centering
    \includegraphics[width=0.68\textwidth]{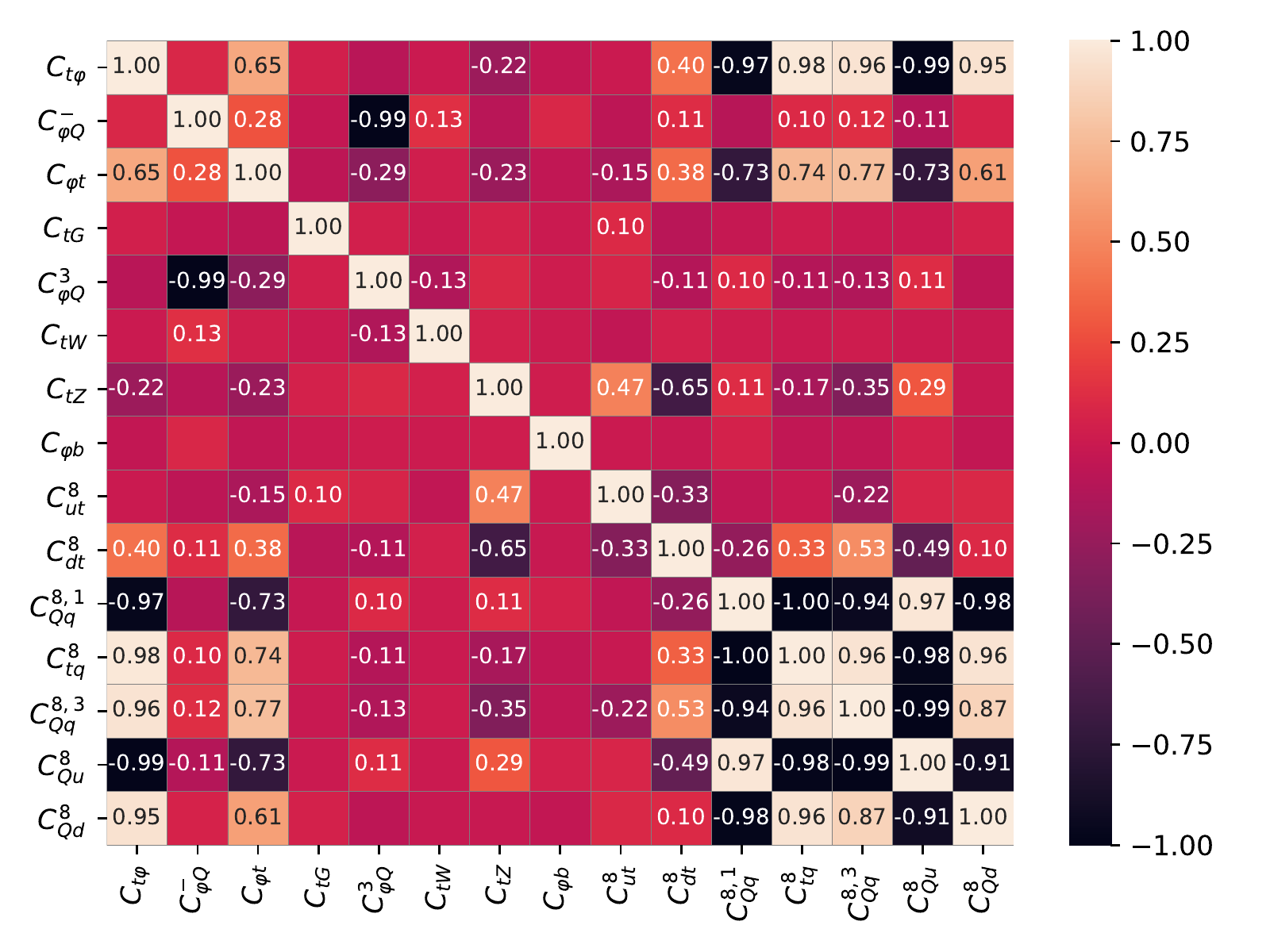}
    \caption{Correlation matrix between the different EFT operators obtained in the linear $(\Lambda^{-2})$ ``4F" robustness test fit described in Section~\ref{sec:limitations}. Cells are filled if the correlation is higher than $10\%$ in absolute value. The operator $O_{\varphi b}$, that modifies only the bottom quark electro-weak couplings, is taken into account in the fit but limits on its coefficients are not reported since the obtained values are not competitive using only the observables considered in our fit.}
    \label{fig:out_base_lin_4f}
\end{figure}


\clearpage

\clearpage
\bibliographystyle{JHEP}
\bibliography{biblio.bib}

\end{document}